\documentclass[11pt,fleqn]{article}

\usepackage{amssymb}
\usepackage{amsmath}
\usepackage{natbib}
\usepackage{vmargin,setspace,verbatim}
\usepackage{graphicx,graphics,color,tikz,tikz-network}
\usepackage{microtype}

\usepackage{libertine}
\usepackage[libertine]{newtxmath}

\definecolor{darkgreen}{rgb}{0,0.2,0}
\definecolor{darkred}{rgb}{0.3,0,0}

\usepackage[colorlinks=true, pdfstartview=FitV, linkcolor=darkgreen, citecolor=darkred, urlcolor=black]{hyperref}

\newcounter{llst}
\newenvironment{abet}{\begin{list}{\rm (\alph{llst})}{\usecounter{llst}
\setlength{\itemindent}{0em} \setlength{\leftmargin}{3em}
\setlength{\labelwidth}{2em} \setlength{\labelsep}{1em}}}{\end{list}}
\newenvironment{numm}{\begin{list}{\rm (\roman{llst})}{\usecounter{llst}
\setlength{\itemindent}{0em} \setlength{\leftmargin}{3.5em}
\setlength{\labelwidth}{2.5em} \setlength{\labelsep}{1em}}}{\end{list}}

\newcounter{axiomatiser}
\newenvironment{axiomatization}{\stepcounter{axiomatiser} \vspace*{2ex} \noindent \textbf{Axiomatization \Roman{axiomatiser}}  \newline \it}{\vspace*{2ex}}

\newcounter{myclaimcount}
\newenvironment{claim}{\stepcounter{myclaimcount} \vspace*{1ex} \noindent \textbf{Claim \Alph{myclaimcount}: } \it}{\vspace*{1ex}}

\newtheorem{theorem}{Theorem}[section]

\newtheorem{definition}[theorem]{Definition}
\newtheorem{expl}[theorem]{Example}

\newtheorem{lemma}[theorem]{Lemma}

\newtheorem{proposition}[theorem]{Proposition}

\newenvironment{proof}[1][Proof]{\noindent \textbf{#1.} }{\hfill
\rule{0.5em}{0.5em}}

\newenvironment{example}{\begin{expl} \rm}{\hfill $\blacklozenge$
\end{expl}}

\setpapersize{A4}
\setmarginsrb{80pt}{40pt}{80pt}{60pt}{12pt}{10pt}{12pt}{30pt}

\begin{document}

\title{\textbf{Expected Values for Variable Network Games}\thanks{This paper was presented at SING16, the virtual European game theory conference in 2021. We thank the SING16 participants for their valuable feedback on our paper. We also thank two anonymous referees for their valuable feedback on a previous draft of this paper.}}

\author{Subhadip Chakrabarti\thanks{Queen's Management School, Queen's University Belfast, 185 Stranmillis Road, Belfast, BT9 5EE, United Kingdom, E-mail: s.chakrabarti@qub.ac.uk} \and
Loyimee Gogoi\thanks{Department of Technology, Operations and Decisions Sciences, Amrut Mody School of Management, Ahmedabad University, India, E-mail: loyimee.gogoi@ahduni.edu.in} \and 
Robert P.~Gilles\thanks{\textbf{Corresponding author} -- Queen's Management School, Queen's University Belfast, 185 Stranmillis Road, Belfast, BT9 5EE, United Kingdom, E-mail: r.gilles@qub.ac.uk} \and
Surajit Borkotokey\thanks{Department of Mathematics, Dibrugarh University, Dibrugarh 786004, Assam, India, Email: surajitbor@gmail.com} \and 
Rajnish Kumar\thanks{Queen's Management School, Queen's University Belfast, 185 Stranmillis Road, Belfast, BT9 5EE, United Kingdom, E-mail: rajnish.kumar@qub.ac.uk}}

\date{September 2021 \\ Revised: October 2022}

\maketitle

\begin{abstract}
\singlespace
\noindent
A network game assigns a level of collectively generated wealth to every network that can form on a given set of players. A \emph{variable network game} combines a network game with a network formation probability distribution, describing certain restrictions on network formation. Expected levels of collectively generated wealth and expected individual payoffs can be formulated in this setting. 

We investigate properties of the resulting expected wealth levels as well as the expected variants of well-established network game values as allocation rules that assign to every variable network game a payoff to the players in a variable network game. We establish two axiomatizations of the Expected Myerson Value, originally formulated and proven on the class of communication situations, based on the well-established component balance, equal bargaining power and balanced contributions properties. Furthermore, we extend an established axiomatization of the Position Value based on the balanced link contribution property to the Expected Position Value.
\end{abstract}

\begin{description}
\singlespace
\item[Keywords:] Network game; variable network game; network formation probabilities; Expected Myerson Value; Expected Position Value; axiomatization.

\item[JEL classification:] C71, D85.
\end{description}

\vfill

\thispagestyle{empty}

\pagebreak 

\setcounter{page}{1} \pagenumbering{arabic}

\section{Introduction}

The understanding of the effects of collaboration and communication through networks on collective wealth generation dates back to \citet{Myerson1977,Myerson1980}. Myerson considered networks to be communication structures that impose constraints on coalition formation in a cooperative game with transferable utilities: A coalition can form if it is connected in the prevailing communication network. This framework---to understand networks as constraints on coalition formation in a cooperative TU-game---is known as a \emph{communication situation}. 

The Shapley Value of the restriction of a cooperative game endowed with a communication network is called the \emph{Myerson Value}, which was seminally introduced by \citet{Myerson1977}. In a communication situation, the precise architecture of the communication network is not important; the resulting class of feasible or formable coalitions determines the resulting values. Two different networks inducing the same partition of the player set will yield an identical restricted game and hence an identical Myerson Value. 

\citet{JacksonWolinsky1996} introduced games in which value stems directly from the network rather than a coalition of players. Such a construct is referred to as a \emph{network game}. In this approach the interaction patterns among players in a network are wealth creating, rather than constraining wealth creation.  Allocation rules for network games specify how the value created by the network game is divided among players. Hence, allocation rules for communication situations can be extended to this setting. Thus, allocation rules in the fixed network setting include the Myerson Value due to \citet{JacksonWolinsky1996} and Position Value due to \citet{Slikker2007}.\footnote{\citet{Jackson2005} argues that the fixed allocation rules are not appropriate in network game settings where the network evolves from players forming and deleting links, and hence came up with the flexible allocation rules.}

The (deterministic) Position Value was seminally introduced by \citet{Meessen1988} as an alternative allocation rule for communication situations and subsequently further developed by \citet{BOT1992}. In this approach links rather than players are considered as the source of all generated wealth. As such, all generated wealth should therefore be allocated to these links. This transforms a communication situation into a link game in which communication links act as players. The Shapley value of the link game now assigns fair values to all links in the network based on the generated wealth. The Position Value of a communication situation is now the distribution of the assigned Shapley link values to all constituting players of these links. \citet{Slikker2005} characterizes the Position Value for communication situations by component balance and the balanced link contributions property.

\citet{Slikker2007} extended the Position Value to the class of network games and characterized this extension using an appropriate formulation of the balanced link contributions property. Slikker's characterization of the Position Value makes it fully compatible with his characterization of the Myerson Value on the same class of network games. In subsequent work, this comparative characterization has been pursued for other classes of cooperative wealth generation.

Games with probabilistic networks were first considered by \citet{Calvo1999} in the context of communication situations. In this framework, links between players are formed stochastically independently according to given probabilities, leading to probabilistic constraints on coalition formation. The resulting network formation probabilities are fully determined by the given link formation probabilities. In particular, the network values are formulated as probabilistic formulations, following the multilinear extension first proposed by \citet{Owen1972}. The wealth created through a coalition is replaced by the expected value based on network-restricted games created by all possible networks that might form on that coalition. \citet{Calvo1999}  extend the Myerson Value to the class of these probabilistic communication situations.

\citet{ProbValue2021} extend the probabilistic perspective on link formation of \citet{Calvo1999} to the realm of network games. This results in \emph{probabilistic network games}, where networks are formed stochastically independently based on given link formation probabilities. \citet{ProbValue2021} extend and characterize the Myerson Value as well as the Position Value to this framework of probabilistic network games.

It can be argued that the independence hypothesis of link formation is questionable. In their paper, Borkotokey et al.~(2021) give a practical example of probabilistic network games over the airline code sharing networks. Passengers travelling on intercontinental flights often using multiple airlines who have code sharing agreements with one another. The passenger pays an up-front fee which is divided among the relevant code sharing airlines in some fashion. But given the competition among airlines, these links are often unstable with airlines terminating existing agreements and forming new agreements. Further, the independence assumption of link formation seems unrealistic in this setting because airlines are looking at the overall strategic situation rather than considering a bilateral agreement in isolation. So, a more general framework of probabilistic network formation is applicable.

This has been developed by \citet{Gomez2008}. In particular, Gomez et al.~(2008) consider a generalization of Calvo et al.~(1999) where the assumption of independent link formation in communication situations is dispensed with. Instead, one assumes an arbitrary probability distribution on the set of all possible networks on a given player set. Gomez et al.~(2008) refer to this as a \emph{generalized probabilistic communication situation}. They extend and characterize the Myerson Value to this more general setting. \citet{Ghintran2012} define and characterize the Position Value in this framework of generalized probabilistic communication situations.

\paragraph{Our framework: Variable network games}

The purpose of our paper is to extend the framework of generalized probabilistic communication situations to the realm of network games. We introduce the notion of a \emph{variable network game} as a combination of a network game and an arbitrary network formation probability distribution. The network game assigns a generated wealth level to every network, while the network formation probability distribution assigns a probability to each network of forming. This framework captures the various frameworks considered in the previous discussion as special cases.

In variable network games, one has to consider the expected levels of wealth that are generated among the players through the networks that can form through which they conduct their affairs. We show that certain properties of the underlying network game are retained in the assignment of expected wealth levels created through these probabilistic networks. 

This allows us to consider allocation rules on the class of variable network games that are founded on familiar allocation rules on the smaller class of network games. Indeed, using expectations of payoffs of these familiar allocation rules over all networks that can form, we arrive at the allocation of the expected wealth that is created in the given variable network game. This allows us to extend the Myerson Value as well as the Position Value to the class of variable network games as the expected payoff allocation rule. This is referred to as the \emph{Expected Myerson Value} and the \emph{Expected Position Value} in the context of our setting.

The Myerson Value on the smaller class of network games has been characterized through two main axiomatizations. The first axiomatization is the one based on the properties of component balance---or component efficiency---and equal bargaining power---or ``fairness''---as seminally conceived by \citet{Myerson1977}. The equal bargaining power property postulates that the change of the allocated payoff is exactly the same for two players if the link between them is removed from the network. In our framework of variable network games this refers to the link being member of a network with zero probability. \citet{JacksonWolinsky1996} extended this axiomatization to the setting of network games. In the current paper we extend this axiomatization further to the Expected Myerson Value on the class of variable network games. 

Our second axiomatization of the Expected Myerson Value is founded on the component balance property in combination with the balanced contributions property. This axiomatization was seminally formulated for the Myerson Value on the class of network games by \citet{Slikker2007}. In our setting the balanced contributions property refers to the effects of any player being ``removed'' in the sense of having zero probability of being a member of a formable network. In particular, the effects on the expected payoffs are equal if other players are ``removed'' from these probabilistic networks. 

Our third axiomatization concerns the extension of Slikker's axiomatization of the Position Value on the class of network games \citep{Slikker2007} to our realm of variable network games. This is founded on the formulation of the balanced link contributions property to the class of variable network games. This property imposes that the accumulated effects of the removal of links in a player's neighborhood is same for all pairs of players. This axiomatization compares directly with the second axiomatization of the Expected Myerson Value and allows an assessment of these two main values for variable network games.

It is clear that further development of the framework of variable network games is warranted and valuable. It allows for the introduction of more tools to properly represent the features of certain interaction situations. We explore here the case of an intermediated trade, where institutional features of the trade situation are represented by the network formation probability distribution rather than the network payoff function. 

\paragraph{Structure of the paper}

In Section 2 we develop the foundations of our approach and introduce the main formal conception of a variable network game. Section 3 introduces the formal conception of a variable network game and explores some properties of the expected wealth that is generated in these variable network games. We also introduce allocation rules on the class of variable network games and define the Expected Myerson and Position Values. Section 4 presents the two main axiomatizations of the Expected Myerson Value and their proofs. Section 5 presents a comparable axiomatization of the Expected Position Value. Section 6 concludes.

\section{Preliminaries: Network games}

Throughout we let $N=\{1,2,\ldots ,n\}$ be a fixed, finite set of players. With a slight abuse of notation, for every set of players $S \subseteq N$, we denote for any player $i \notin S$ the expanded set $S+i = S \cup \{ i \}$ and for any player $i \in S$ the reduced set $S-i = S \setminus \{ i \}$. In particular, the set $N-i$ for $i \in N$ is the set of all players other than $i$. Furthermore, we denote by $\# S$ the number of elements in a set $S$. 

\subsection{Cooperative games and the Shapley value}

A \emph{cooperative game} on a player set $N$ is a mapping $\omega \colon 2^N \to \mathbb R$ such that $\omega (\varnothing )=0$. A cooperative game assigns to every non-empty coalition of players $S \subseteq N$ some ``worth'' $\omega (S)$, representing a collectively generated wealth. The function $\omega$ is also referred to as a \emph{characteristic function}.

An \emph{allocation rule} is a mapping that assigns to every cooperative game $\omega$ some vector $x \in \mathbb R^N$. \citet{Shapley1953} seminally introduced the seminal allocation rule $\phi$ that assigns to every cooperative game $\omega$ an allocation $\phi (\omega ) \in \mathbb R^N$ given for every player $i \in N$ by
\begin{equation}
	\phi_i (\omega ) = \sum_{S \subseteq N-i} \, \frac{\#S! \, (n- \# S -1)!}{n!} \,  [ \, \omega (S+i) - \omega (S) \, ]
\end{equation}
Shapley showed in his seminal paper that the Shapley value is the unique allocation rule that satisfies the following four properties:\footnote{We remark that these properties are formalised for the Shapley value $\phi$ rather than a more general allocation rule. }
\begin{description}
	\item[Efficiency:] For every $\omega$ it holds that $\sum_{i \in N} \phi_i ( \omega ) = \omega (N)$;
	\item[Null-player property:] If $i \in N$ is a null player in $\omega$ in the sense that $\omega (S) = \omega (S-i)$ for all $S \subseteq N$, then $\phi_i (\omega )=0$;
	\item[Symmetry:] All symmetric players $i,j \in N$ such that $\omega (S +i ) = \omega (S+j)$ for all $S \subseteq N \setminus \{ i,j \}$ are treated equally, in the sense that $\phi_i ( \omega ) = \phi_j (\omega )$,  and;
	\item[Linearity:] For all cooperative games $\omega , \omega'$ it holds that $\phi ( \alpha \omega + \beta \omega' ) = \alpha \phi (\omega ) + \beta \phi (\omega' )$ for all $\alpha , \beta \in \mathbb R$.
\end{description}
The Shapley value lies at the foundation of the wealth allocation rules for network games and variable network games considered in this paper.

\subsection{Network preliminaries}

Given the player set $N $, a \emph{link} between two distinct players $i \in N$ and $j \in N$ is defined as the binary set $ij = \{ i,j \}$, representing an undirected relationship between $i$ and $j$.\footnote{In the following discussion and throughout the remainder of this paper, we use and extend the notation seminally introduced by \citet{JacksonWolinsky1996}.}  Clearly, $ij$ is equivalent to $ji$. The set of all possible links on $N$ is denoted by $g_N = \{ ij \mid i,j \in N \mbox{ and } i \neq j \}$.

A \emph{network} on $N$ is any set of links $g \subseteq g_N$.\footnote{This definition implies that networks are \emph{simple graphs} in the sense of standard mathematical graph theory.} In particular, $g_N$ is called the \emph{complete} network, while $g_0 = \varnothing$ is the \emph{empty} network. The class of all possible networks on $N$ is given by $\mathbb G^N = \{ g \mid g \subseteq g_N \}$. Clearly, $\# g_{N}=\frac{n(n-1)}{2}$ and $\# \mathbb G^N = 2^{\frac{n(n-1)}{2}}$.

For every network $g\in \mathbb{G}^{N}$ and every player $i\in N$ we denote $i$'s \emph{neighborhood} in $g$ by $N_{i}(g)=\{j\in N\mid j\neq i$ and $ij\in g\}$ and her \emph{link neighborhood} by $L_{i}(g)=\{ij\in g\mid j\in N_{i}(g)\}\subseteq g$. Furthermore, $n_{i}(g)=\#N_{i}(g) = \# L_i (g)$. In particular, we introduce $L_i = L_i (g_N)$ as the set of all potential links in which player $i$ participates.

We also define $N(g)=\cup _{i\in N}N_{i}(g)$ and let $n(g)=\#N(g)$ with the convention that if $N(g)=\emptyset $, we let $n(g)=1$.\footnote{We emphasize here that if $N(g)\neq \emptyset $, we have that $n(g)\geqslant 2$. Namely, in those cases the network has to consist of at least one link.} $n(g)$ will be referred to as \emph{size} of the network $g$.

We say that player $i \in N$ is an \emph{isolated player} in network $g \in \mathbb G^N$ if $i \notin N(g)$. This implies that for isolated player $i$ in $g$ it holds that $N_i(g) = \varnothing$. We denote the set of all isolated players in network $g$ by $N_0 (g) = N \setminus N(g)$.

\paragraph{Connectivity in networks}

A \emph{path} connecting $i \in N$ and $j \in N$ with $i \neq j$ in network $g \in \mathbb G^N$ is an ordered set of distinct players $P_{ij} = \{i_{1},i_{2},\ldots ,i_{p}\}\subseteq N(g)$ with $p\geqslant 2$ such that $i_{1}=i$, $i_{p}=j$, and $i_k i_{k+1} \in g$ for all $k = 1, \ldots ,p-1$. We say that $i,j \in N$ with $i \neq j$ are \emph{connected} in $g$ if there exists a path $P_{ij} \subseteq N(g)$ between $i$ and $j$ and \emph{disconnected} otherwise. A network is \emph{connected} if all pairs of players $i,j \in N$ with $i \neq j$ are connected. In particular, for a connected network $g$ it holds $N(g)=N$ and $N_0 (g) = \varnothing$.

The network $h \in \mathbb G^N$ is a \emph{component} of $g \in \mathbb G^N$ if $h \subseteq g$ is connected and for any $i \in N(h)$ and $j\in N(g)$, $ij\in g$ implies $ij\in h$. In other words, a component is a maximally connected subnetwork of $g$. We denote the set of network components of the network $g$ by $C(g)$. Note that for any connected network $g \in \mathbb G^N$ it holds that $C(g) = \{ g \}$. In particular, $C(g_N) = \{ g_N \}$ and $C(g_0) = \varnothing$.

\paragraph{Restrictions of networks}

Let $g \in \mathbb G^N$ and let $S \subseteq N$ be some set of players. The set of all links within the coalition $S$ can be represented as $g_S = \{ ij \mid i,j \in S$ and $i \neq j \}$. Now, the \emph{restriction} of $g$ to $S$ is the network defined as $g|S = \{ ij \in g \mid i,j \in S$ and $i \neq j \} = g \cap g_S$. The restricted network $g|S$ is obviously a subnetwork of $g$.

For a subnetwork $h \subseteq g$, we denote by $g-h = g \setminus h$ the network that results after the removal from $g$ of all links in the subnetwork $h$. Similarly, for any network $h \subseteq g_N-g = g_N \setminus g$ we denote by $g+h = g \cup h$ the network that results from $g$ after adding all links in $h$ to the network $g$.\footnote{Note here that by selection of $h \subseteq g_N \setminus g$ it holds that $g \cap h = \varnothing$.} Clearly, for $h \subseteq g$ it holds that $(g-h)+h = g$.

\subsection{Network formation probability distributions}

Following \citet{Gomez2008} we investigate the formal description of the probabilistic emergence or formation of networks on a given set of players. These probabilistic structures are introduced as additional modelling tools to understand certain phenomena in wealth creation processes observed in the economy. In particular, these probabilistic structures can be used to describe basic link formation failures or fuzziness related to the formation of a network. 

Before discussing some motivating examples, we formally introduce the notion of a network formation probability distribution that assigns to every network a probability that it forms.
\begin{definition}
	A \textbf{network formation probability distribution} on $N$ is a map $\rho \colon \mathbb G^N \to [0,1]$ such that $\sum_{g \in \mathbb G^N} \rho (g) =1$.
	\\
	The class of all network formation probability distributions on $N$ is denoted by 
	\begin{equation}
		\mathbb P^N = \left\{ \left. \rho \in [0,1]^{\# \mathbb G^N} \, \right| \, \mbox{$\sum_{g \in \mathbb G^N}$} \, \rho (g) =1 \, \right\} .
	\end{equation}
	Note that $\dim \mathbb P^N = \# \mathbb G^N -1 = 2^{\frac{1}{2} n(n-1)} -1$.
\end{definition}
The notion of a network formation probability distribution was introduced by \citet{Gomez2008}, generalising the link-based network formation approach of \citet{Calvo1999}. A network formation probability distribution naturally results in the following conceptions.
\begin{definition}
	Let $\rho \in \mathbb P^N$ be some network formation probability distribution on $N$.
	\begin{numm}
		\item A network $g \in \mathbb G^N$ is \textbf{formable} under $\rho$ if $\rho (g) >0$. The class of formable networks under $\rho$ is given by $\, \mathbb G( \rho ) = \{ g \in \mathbb G^N \mid \rho (g)>0 \}$, which can also be denoted as the \textbf{support} of $\rho$.
		\item The \textbf{extent} of the distribution $\rho$ is the network $g ( \rho ) \in \mathbb G^N$ defined as the union of all formable networks, namely, $g( \rho ) = \cup \, \mathbb G( \rho ) \in \mathbb G^N$.
		\item A network $h \in \mathbb G^N$ is a \textbf{component} of $\rho$ if $h \in C( g ( \rho ))$ is a component of the extent $g ( \rho )$ and a player $i \in N$ is \textbf{isolated} in $\rho$ if $i \in N_0 (g( \rho ))$ is an isolated player in the extent $g  ( \rho )$.
	\end{numm}
\end{definition}
Formable networks are those that are assigned a positive formation probability. The extent of the network formation distribution is simply the collection of all links that are part of a formable network. Hence, a link is not in the extent if it is not part of any formable network and, as such, will form with zero probability. Therefore, a player is isolated if there is a zero probability that she is linked to any other player under the given network formation probability distribution.

The extent of a network formation probability distribution is recognised as the network that consists of all links that form with positive probabilities, extending the definition of the support developed in \citet{ProbValue2021} for probabilistic networks to the class $\mathbb P^N$. The components and isolated players in the extent of a network formation probability distribution $\rho$ critically determine the main features of those networks that can emerge under $\rho$.

\paragraph{A motivating example: Intermediated bilateral trade}

The following examples provide some motivation for the study of these structures. We argue that these mathematical devices can represent institutional aspects of network formation into the framework of network games. These examples are mainly motivated by the discussion in \citet[Section 2.1]{Calvo1999} of a simple bilateral trading case with potential intermediation. There the case is developed from a link-based probabilistic perspective. We investigate that here, prior to generalizing to an implementation of network-based probabilities representing an institutional constraint on network formation.

\medskip
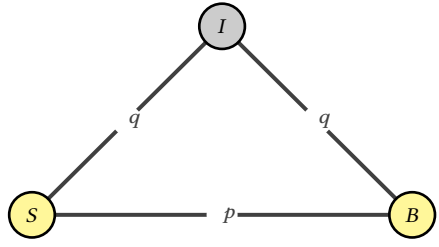
\begin{figure}[t]
	\begin{center}
	\begin{tikzpicture}[scale=0.5]
		\Vertex[x=0,y=0,label=$S$,color=yellow!50]{1}
		\Vertex[x=10,y=0,label=$B$,color=yellow!50]{2}
		\Vertex[x=5,y=5,label=$I$,color=black!20]{3}
		\Edge[label=$q$](1)(3)
		\Edge[label=$q$](2)(3)
		\Edge[label=$p$](1)(2)
	\end{tikzpicture}
	\end{center}
	\caption{Network of an intermediated buyer-seller situation}
\end{figure}

\begin{example} \label{Motiv-A} \textbf{(Independent link formation)} \\
	Consider a trading situation with one seller $S$ and one buyer $B$ who can generate mutual gains from trade by the transfer of an object. The total wealth that is created is set to one (1). The trade can either be executed directly between $S$ and $B$ or through the intermediation of an intermediary $I$. The player set can be identified as $N = \{ S,B,I \}$.
	\\
	In the standard approach to network games, the links between these three players either exist or not. \citet{Myerson1977,Myerson1980} seminally investigated these situations. Under the Myerson hypotheses, the full wealth of 1 can be realised in networks $g \in \mathbb G^N$ such that $SB \in g$ and/or $\{ SI,BI \} \subseteq g$. In all other networks the generated wealth is zero, since no trade can be accomplished. We can describe this by an appropriately constructed network formation probability distribution on $\mathbb G^N$. 
	\\
	\citet{Calvo1999} introduced the instrument of probabilistic links to describe that relationships are formed subject to certain conditions. Hence, probabilities on the links in $g_N$ are introduced, representing the fundamental uncertainty that links can be formed. Assuming that both $SI$ and $BI$ are formed with equal probability $q \in [0,1]$ and that the critical link $SB$ is formed with probability $p \in [0,1]$, we arrive at a graphical depiction of the communication situation as depicted in Figure 1.
	\\
	Next we introduce the hypothesis that all links are formed \emph{independently}. This implies that a network forms with a probability that is determined as a product of link formation probabilities or their non-formation. We can now determine the generated network formation probability distribution as shown in Table 1.
	
	\begin{table}[h!]
	\begin{center}
	\begin{tabular}{|c|c|}
		\hline
		Network & Probability \\ \hline
		 $\varnothing$ & $(1-p)(1-q)^2$ \\
		 $\{ SI \}$ & $(1-p)(1-q)q$ \\
		 $\{ BI \}$ & $(1-p)(1-q)q$ \\
		 $\{ SB \}$ & $p (1-q)^2$ \\
		 $\{ SI,BI \}$ & $(1-p)q^2$ \\
		 $\{ SI,SB \}$ & $p(1-q)q$ \\
		 $\{ BI,SB \}$ & $p(1-q)q$ \\
		 $g_N$ & $p q^2$ \\
		\hline
	\end{tabular}
	\end{center}
	\caption{Network formation probabilities in Example \ref{Motiv-A}.}
	\end{table}
	
	\noindent
	The expected generated wealth can now be computed as
	\begin{align*}
		\mathbb E (W) & = \mathrm{prob} \, \{ SB \in g \} +  \mathrm{prob} \, \{ \, \{ SI,BI \} \subseteq g \} -  \mathrm{prob} \, (g_N) \\
		& = p + q^2 - pq^2 = p + (1-p) q^2
	\end{align*}
	Note that $\mathbb E (W) =1$ if $p=1$ and/or $q=1$. \citet{Calvo1999} only considered the case of $q=1$ in their motivating discussion.
\end{example}

\noindent
Example \ref{Motiv-A} discusses the special case of network formation being based on the independent formation of the individual links that make up the network---further explored in \citet{Calvo1999} and \citet{ProbValue2021}. In particular, let $p \colon g_N \to [0,1]$ be some assignment of link formation probabilities on $g_N$. Under independence of link formation, the probability that a network $g \in \mathbb G^N$ forms is then given by the multilinear form
\begin{equation}
	\mu_p (g) = \prod_{ij \in g} p_{ij} \cdot \prod_{ij \notin g} \left( 1 - p_{ij} \right) .
\end{equation}
It holds that $\mu_p \in \mathbb P^N$ for every link formation probability assignment $p \colon g_N \to [0,1]$.

The next example discusses a case in which the network formation probability distribution is based on a link formation probability distribution representing an institutional feature in the network formation process.

\begin{example} \label{Motiv-B} \textbf{(Institutional network formation)} \\
	Consider again the intermediated bilateral trade situation discussed in Example \ref{Motiv-A}. We amend this case by assuming that this trade occurs in an institutional framework of a certain implementation of contract law and that a relationship can only be effectuated if notarised. We assume further that the intermediary $I$ is a notary and that trade can, therefore, only occur in a network to which $I$ is connected.\footnote{Therefore, we de facto assume that either $S$ or $B$ or both of these players need to know the intermediary $I$ to have access to her notary services.} Hence, trade can only be executed in networks $g \in \mathbb G^N$ such that $B$ and $S$ can trade \emph{as well as} $I \in N(g)$. 
	\\
	Furthermore, we assume that the costs related to the formal notarisation of the contract by $I$ is negligible in relation to the wealth created through the trade between $S$ and $B$.
	\\
	We apply again the assumption that network formation is founded on the link formation probabilities of $q \in [0,1]$ of $SI$ and $BI$ forming and of $p \in [0,1]$ of $SB$ forming. However, now networks can only be formed if $I$ is connected, which probability is given by $q(2-q)$. This results in conditional probabilities of network formation in relation to the probabilities reported in Example \ref{Motiv-A}. These conditional probabilities are given in Table 2 below.
	
	\begin{table}[h!]
	\begin{center}
	\begin{tabular}{|c|c|}
		\hline
		Network & Probability \\ \hline
		 $\varnothing$ & $0$ \\
		 $\{ SI \}$ & $\frac{(1-p)(1-q)}{2-q}$ \\
		 $\{ BI \}$ & $\frac{(1-p)(1-q)}{2-q}$ \\
		 $\{ SB \}$ & $0$ \\
		 $\{ SI,BI \}$ & $\frac{(1-p)q}{2-q}$ \\
		 $\{ SI,SB \}$ & $\frac{p(1-q)}{2-q}$ \\
		 $\{ BI,SB \}$ & $\frac{p(1-q)}{2-q}$ \\
		 $g_N$ & $\frac{p q}{2-q}$ \\
		\hline
	\end{tabular}
	\end{center}
		\caption{Network formation probabilities in Example \ref{Motiv-B}.}
	\end{table}

	\noindent
	The expected generated wealth in this institutional network formation setting can be computed as
	\begin{align*}
		\mathbb E (W') & = \mathrm{prob} \, \{ SB \in g \} +  \mathrm{prob} \, \{ \, \{ SI,BI \} \subseteq g \} -  \mathrm{prob} \, (g_N) \\
		& =  p +\tfrac{q}{2-q} - \tfrac{pq}{2-q} = \tfrac{q+2p(1-q)}{2-q}
	\end{align*}
	Note that as before $\mathbb E (W') =1$ if $p=1$ and/or $q=1$. 
	\\
	Moreover, we easily compute that $\mathbb E (W') > \mathbb E(W)$ if and only if $(p-1)(q-1)^2 <0$ if and only if $p<1$ as well as $q<1$. This implies that this simple application shows that institutional embedding increases the expected generated wealth in this simple bilateral trade situation. 
\end{example}

\paragraph{Restrictions of network formation probability distributions}

Next we discuss the notion of restricting a network formation probability distribution to a certain given (deterministic) network. This implies that its extent is limited to the given network. The following definition formalises this by transferring network formation probabilities of networks extending beyond the imposed restriction to the networks that form within the imposed restricted extent. The applied conception is due to \citet[page 543]{Gomez2008}.
\begin{definition} \label{def:rho-Restriction}
	Let $\rho \in \mathbb P^N$ be some network formation probability distribution and let $g \in \mathbb G^N$ be some given network on $N$. Then the \textbf{restriction of $\rho$ to $g$} is the modified network formation probability distribution $\rho_g \in \mathbb P^N$ defined by 
	\begin{equation}
		\rho_g (h) = \left\{
		\begin{array}{cl}
			\sum_{h' \subseteq g_N \setminus g} \, \rho (h \cup h') & \mbox{if } h \subseteq g	 \\[1ex]
			0 & \mbox{otherwise}
		\end{array}
		\right.
	\end{equation}
\end{definition}
In this definition the formation probabilities of subnetworks of $g_N \setminus g$ are transferred to subnetworks of $g$ itself. It should be clear that the extent of the restriction of $\rho$ to some network $g$ is determined as $g(\rho_g) = g(\rho ) \cap g$.

As special cases of this notion of a restriction of a network formation probability distribution to a given network, we introduce devices for removing individual links and players. For the computation of these restrictions we can state the following proposition.
\begin{proposition} \label{prop:rho-restrict}
	Let $\rho \in \mathbb P^N$ and let $i,j \in N$ with $i \neq j$.
	\begin{abet}
		\item Let $\rho^{-ij} = \rho_{g_N-ij}$ be the restriction of $\rho$ to $g_N-ij$. Then for every network $g \in \mathbb G^N \colon$
			\begin{equation} \label{eq:rho-restrictA}
				\rho^{-ij} (g) = \left\{
				\begin{array}{ll}
					\rho (g) + \rho (g+ij) & \mbox{if } \quad ij \notin g \\
					0 & \mbox{if } \quad ij \in g
				\end{array} 
				\right. .
			\end{equation}
			Furthermore, it holds that $g \left( \rho^{-ij} \right) = g ( \rho )-ij$ and
			\begin{equation} \label{eq:G-rho-ij}
				\mathbb G \left( \rho^{-ij} \right) = \left\{ g \in \mathbb G^N \mid ij \notin g \  \mbox{ and } \  \mathbb G ( \rho ) \cap \{ g, g+ij \} \neq \varnothing \, \right\}
			\end{equation}
		\item Let $\rho^{-i} = \rho_{g_{N-i}}$ be the restriction of $\rho$ to $g_{N-i} = g_N \setminus L_i (g_N)$. Then for every network $g \subseteq g_{N-i}$ it holds that
			\begin{equation} \label{eq:rho-restrictB}
				\rho^{-i} (g) = \rho (g) + \sum_{\varnothing \neq h \subseteq L_i (g (\rho ))} \rho (g \cup h)
			\end{equation}
			and for any $g \not\subseteq g_{N-i} \colon \rho^{-i} (g) = 0$. \\
			Furthermore, it holds that $g \left( \rho^{-i} \right) = g ( \rho ) \setminus L_i (g_N)$ and
			\begin{equation} \label{eq:G-rho-i}
				\mathbb G \left( \rho^{-i} \right) = \left\{ g \in \mathbb G^N \mid g \cap L_i (g_N) = \varnothing \ \ \mbox{and} \ \ \mathbb G (\rho ) \cap \{ g+h \mid h \subseteq L_i (g_N) \} \neq \varnothing \, \right\}
			\end{equation}
	\end{abet}
\end{proposition}
\begin{proof}
	It is clear that (\ref{eq:rho-restrictA}) in assertion (a) follows immediately from the definition of $\rho_{G_N-ij}$ as introduced in Definition \ref{def:rho-Restriction}. \\
	Next, note that $g \left( \rho^{-ij} \right) = g ( \rho )-ij$ follows immediately from (\ref{eq:rho-restrictA}). \\
	To prove (\ref{eq:G-rho-ij}), take $g \in \mathbb G \left( \rho^{-ij} \right)$. Then by (\ref{eq:rho-restrictA}) it follows that $ij \notin g$. \\ Next, take any $g \in \mathbb G^N$ with $ij \notin g$. Then by (\ref{eq:rho-restrictA}), $\rho^{-ij} (g) = \rho (g) + \rho (g+ij) >0$ if and only if $\rho (g) >0$ and/or $\rho\ (g+ij)>0$ if and only if $\{ g, g+ij \} \cap \mathbb G (\rho ) \neq \varnothing$. \\ Finally, take any $g \in \mathbb G^N$ with $ij \in g$. Then $\rho^{-ij} (g)=0$, implying that $g \notin \mathbb G \left( \rho^{-ij} \right)$. \\ This show that (\ref{eq:G-rho-ij}) indeed holds.
	\\[1ex]
	To show (\ref{eq:rho-restrictB}) in assertion (b), let $\rho \in \mathbb P^N$ and $i \in N$. From Definition \ref{def:rho-Restriction} it is obvious that for every $g \in \mathbb G^N$ with $g \not\subseteq g_{N-i} \colon \rho^{-i} (g) = \rho_{g_{N-i}} (g) = 0$. \\ Next let $g \in \mathbb G^N$ with $g \subseteq g_{N-i}$. Then it holds by definition that
	\[
	\rho^{-i} (g) = \rho_{g_{N-i}} (g) = \sum_{h \subseteq L_i (g_N)} \rho (g \cup h) = \rho (g) + \sum_{\varnothing \neq h \subseteq L_i (g_N)} \rho (g \cup h)
	\]
	Furthermore, from the definitions it is clear that all networks with links outside the extent $g( \rho )$ have zero formation probability under $\rho$. Hence,
	\[
	\rho^{-i} (g) = \rho (g) + \sum_{\varnothing \neq h \subseteq L_i (g_N) \cap g( \rho )} \rho (g \cup h) = \rho (g) + \sum_{\varnothing \neq h \subseteq L_i (g (\rho ))} \rho (g \cup h) .
	\]
	This shows (\ref{eq:rho-restrictB}).
	\\
	To show (\ref{eq:G-rho-i}), take $g \in \mathbb G^N$. Now if $g \cap L_i (g_N) = \varnothing$, $\rho^{-i} (g) = \sum_{h \subseteq L_i (g_N)} \rho (g+h) >0$ if and only if $ \rho (g+h)>0$ for some $h \subseteq L_i (g_N)$ (including $h = \varnothing$) if and only if $\{ g+h \mid h \subseteq L_i (g_N) \} \cap \mathbb G( \rho) \neq \varnothing$.
	\\
	Moreover, if $g \cap L_i (g_N) \neq \varnothing$, it follows immediately that $\rho^{-i} (g) = 0$ and, therefore, $g \notin \mathbb G \left( \rho^{-i} \right)$. \\
	Finally, from (\ref{eq:G-rho-i}) we immediately see that $g \left( \rho^{-i} \right) = g ( \rho ) \setminus L_i (g_N)$. This shows assertion (b) of the proposition.
\end{proof}

\subsection{Network games}

\citet{JacksonWolinsky1996} seminally introduced the notion of a network game on the player set $N$. It is assumed that every cooperation among players intermediated through some configuration of relationships between these players creates a level of wealth that is determined by the architecture of the network that is formed. This leads to the introduction of a function that assigns a wealth level to every network that can be formed on $N$.

Formally, a \emph{network game} on $N$ is a function $v \colon \mathbb G^N \to \mathbb R$ such that $v (g_0) = 0$. This leads to the class of all network games on $N$ to be defined as
\begin{equation}
	\mathbb V^N = \{ v \mid v \colon \mathbb G^N \to \mathbb R \mbox{ such that } v(g_0) =0 \}
\end{equation}
The class $\mathbb V^N$ has been the subject of study of numerous contributions to game theoretic network analysis. It is clear that $\mathbb V^N$ is a $\left( 2^{\tfrac{1}{2} n(n-1)}-1 \right)$-dimensional Euclidean space.

We recall that a network game $v \in \mathbb V^N$ is \emph{component additive} if $v(g) = \sum_{h \in C(g)} v(h)$ for all $g \in \mathbb G^N$. This additional property will be used frequently throughout the next sections on variable network games.

\paragraph{Allocation rules on $\mathbb V^N \times \mathbb G^N$}

Following the notation introduced by Jackson and Wolinsky, an \emph{allocation rule} on the class of network games is a mapping $Y \colon \mathbb V^N \times \mathbb G^N \to \mathbb R^N$ that for every network game $v \in \mathbb V^N$ assigns to every player $i \in N$ in a network $g \in \mathbb G^N$ an allocated value $Y_i (v,g)$ such that $Y_i (v,g) = 0$ for every isolated player $i \in N_0 (g)$.\footnote{The imposed property that isolated players are assigned a zero value is a required hypothesis, since under standard properties such as balance and component balance one can only show that $\sum_{i \in N_0 (g)} Y_i (v,g)=0$. For a further discussion we also refer to \citet{ProbValue2021}.}

Allocation rules on network games can satisfy a number of standard properties that have been introduced and investigated in the literature. We list the most relevant of these properties below.
\begin{itemize}
	\item An allocation rule $Y$ on $ \mathbb V^N \times \mathbb G^N$ is \emph{balanced} if for every network game $v \in \mathbb V^N$ and every network $g \in \mathbb G^N \colon \sum_{i \in N} Y_i (v,g) = v(g)$.
	\item An allocation rule $Y$ on $ \mathbb V^N \times \mathbb G^N$ is \emph{component balanced} if for every component additive network game $v \in \mathbb V^N$, every network $g \in \mathbb G^N$ and all of its components $h \in C(g) \colon \sum_{i \in N(h)} Y_i (v,g) = v(h)$. \\ Component balance implies balance for component additive network games.
	\item An allocation rule $Y$ on $ \mathbb V^N \times \mathbb G^N$ satisfies \emph{equal bargaining power}\footnote{The equal bargaining power property was referred to as the ``fairness'' property by \citet{Myerson1977}.} if for every network game $v \in \mathbb V^N$ and every network $g \in \mathbb G^N$ it holds for all pairs $i,j \in N$ with $ij \in g$ that
	\[
		Y_i (v,g) - Y_i (v,g-ij) = Y_j (v,g) - Y_j (v,g-ij) .
	\]
	\item An allocation rule $Y$ on $ \mathbb V^N \times \mathbb G^N$ satisfies the \emph{balanced contributions property} if for every network game $v \in \mathbb V^N$ and every network $g \in \mathbb G^N$ it holds for all players $i,j \in N$ that
	\[
		Y_i (v,g) - Y_i (v,g \setminus L_j ) = Y_j (v,g) - Y_j (v,g \setminus L_i ) .
	\]
	\item An allocation rule $Y$ on $\mathbb V^N \times \mathbb G^N$ satisfies the \emph{balanced link contributions property} if for every network game $v \in \mathbb V^N$ and every network $g \in \mathbb G^N$ it holds for all players $i,j \in N$ with $i\neq j$ that
	\[
		\sum_{jk\in L_{j}(g)}  \left( Y_{i}(v,g)-Y_{i}(v,g-jk) \, \right) = \sum_{ik\in L_{i}(g)} \left( Y_{j}(v,g)-Y_{j}(v,g-ik) \, \right) .
	\]
\end{itemize}
The properties listed above have been used to characterise the most common allocation rules on the class of network games.

\paragraph{The Myerson Value for network games}

\citet{JacksonWolinsky1996} formulated the Myerson Value as an allocation rule on the class of network games that is an extension of the allocation rule for communication situations seminally introduced by \citet{Myerson1977}. Formally, the \emph{Myerson Value} on the class of network games is the allocation rule $Y^m \colon \mathbb V^N \times \mathbb G^N \to \mathbb R^N$ defined by
\begin{equation} \label{eq:Y-m}
	Y^m_i (v,g) = \sum_{S \subseteq N-i} \, \frac{\#S! \, (n- \# S -1)!}{n!} \,  [ \, v(g|S+i) - v(g|S) \, ]
\end{equation}
\citet{JacksonWolinsky1996} show---as an extension of the main result of \citet{Myerson1977} for communication situations---that the Myerson Value $Y^m$ is the unique allocation rule on the class of network games $\mathbb V^N \times \mathbb G^N$ that satisfies component balance and the equal bargaining power property. Furthermore, \citet{Slikker2007} shows that the Myerson Value $Y^m$ is the unique allocation rule on the class of network games $\mathbb V^N \times \mathbb G^N$ that satisfies component balance and the balanced contributions property.

\citet{JacksonWolinsky1996} also showed that the Myerson Value $Y^m$ on the class of network games $\mathbb V^N \times \mathbb G^N$ is the Shapley Value of an associated cooperative game $\hat\omega_{(v,g)} \colon 2^N \to \mathbb R$ defined for every network game $v \in \mathbb V^N$ and network $g \in \mathbb G^N$ defined by $\hat\omega_{(v,g)} (S) = v(g|S)$ for any coalition $S \subseteq N$. Hence, $Y^m (v,g) = \phi ( \hat\omega_{(v,g)} )$.

\paragraph{The Position Value for network games}

\citet{Slikker2007} introduced the Position Value as an allocation rule on the class of network games by extending the earlier definition of \citet{Meessen1988} to this extended framework. Formally, the \emph{Position Value} on the class of network games is the allocation rule $Y^{p} \colon \mathbb{V}^{N} \times \mathbb{G}^{N} \to \mathbb{R}^{N}$ defined by 
\begin{equation} \label{eq:Y-p}
	Y_{i}^{p}(v,g)=\tfrac{1}{2} \ \sum_ {ij\in g} \ \sum_{h \subseteq g - ij} \frac{\#h ! \, \left( \#g-\#h-1\right) !}{\#g!} \, \left( v(h+ij)-v(h) \right)
\end{equation}
\citet{Slikker2007} shows that the Position Value $Y^p$ is the unique allocation rule on the class of network games $\mathbb V^N \times \mathbb G^N$ that satisfies component balance and the balanced link contributions property. This insight allows a complete comparison between the Myerson and Position Values for network games.

\section{Variable network games}

In this paper we extend the class of network games to the larger class of network wealth creation situations in which network formation processes are assumed to be probabilistic. This is represented by a combination of a network game and a network formation probability distribution.
\begin{definition}
	A \textbf{variable network game} is a pair $(v, \rho ) \in \mathbb V^N \times \mathbb P^N$ consisting of a network game $v$ describing the potential wealth levels created through the formed networks and a network formation probability distribution $\rho$ describing the probabilities with which networks form.
	\\
	The \textbf{expected wealth} that is created in the variable network game $(v, \rho ) \in \mathbb V^N \times \mathbb P^N$ is defined as
	\begin{equation}
		\mathbb E (v, \rho ) = \sum_{g \in \mathbb G^N} \, \rho (g) \cdot v(g)
	\end{equation}
\end{definition}
There are some properties satisfied by the expected wealth that is created in a variable network game. First, we consider the case that a network game $v \in \mathbb V^N$ is component additive. Component additivity reflects explicitly that all collective wealth is generated in the component through which the players interact. Therefore, there are no externalities across disconnected components of a network and the total generated wealth is simply the sum of the wealth generated in the constituting components of the network.

Formally, a network game $v \in \mathbb V^N$ is \emph{component additive} if for every network $g \in \mathbb G^N$ it holds that $v(g) = \sum_{h \in C(g)} v(h)$. Throughout the remainder of the paper, we assume that all network games considered are component additive.
\begin{proposition}
	Let $(v, \rho ) \in \mathbb V^N \times \mathbb P^N$ be such that $v$ is component additive. Then it holds that
	\begin{equation}
		\mathbb E (v, \rho ) = \sum_{h \in C(g( \rho ))} \ \sum_{h' \subseteq h} \rho_h (h') \cdot v(h')
	\end{equation}
\end{proposition}
\begin{proof}
	From the component additivity of $v$ it immediately follows that
	\begin{align*}
		\mathbb E (v, \rho ) & = \sum_{g \subseteq g( \rho )} \rho (g) \cdot v(g) = \sum_{g \subseteq g( \rho )} \rho (g) \cdot \sum_{h \in C(g( \rho ))} v(g \cap h ) \\[1ex]
		& = \sum_{h \in C( g ( \rho ))} \sum_{g \subseteq g(\rho )} \rho (g) \cdot v(g \cap h ) = \sum_{h \in C(g( \rho ))} \sum_{h' \subseteq h} \rho_h (h') \cdot v(h') .
	\end{align*}
	by definition of the restriction $\rho_h$ of $\rho$ to any $h \subseteq g(\rho )$. This shows the assertion.
\end{proof}

\bigskip\noindent
Second, we consider the marginal contributions of individual links and players to the expected wealth that is created by a variable network game. Recall that in this context the deletion of a link $ij \in g_N$ or a player $i \in N$ from a network formation probability distribution $\rho \in \mathbb P^N$ is expressed through the restrictions $\rho^{-ij}$ and $\rho^{-i}$, respectively. The next definitions formalise the contributions made by links and players in regular network games. 

Let $v \in \mathbb V^N$ be a network game and let $g \in \mathbb G^N$ be a given network. The \emph{marginal contribution of a link} $ij \notin g$ to the network game $v$ at network $g$ is given by $\Delta_{ij} (v,g) = v(g+ij) - v(g)$.

Similarly, the \emph{marginal contribution of a player} $i \in N(g)$ to the network game $v$ at network $g$ is given by $\Delta_i (v,g) = v(g) - v(g \setminus L_i)$.

The next proposition collects properties that describe the marginal contributions of links and players to a variable network game.
\begin{proposition} \label{prop-contributing}
	Let $(v, \rho ) \in \mathbb V^N \times \mathbb P^N$ be a variable network game such that $v$ is component additive.
	\begin{abet}
		\item For every link $ij \in g_N$ it holds that
		\begin{equation}
			\mathbb E(v, \rho^{-ij} ) = \mathbb E(v, \rho ) - \sum_{g \colon ij \notin g} \rho (g+ij) \cdot \Delta_{ij} (v,g)
		\end{equation}
		\item For every player $i \in N$ it holds that
		\begin{equation}
			\mathbb E (v, \rho^{-i} ) = \mathbb E(v, \rho ) - \sum_{g \colon i \in N(g)} \rho (g) \cdot \Delta_i (v,g)
		\end{equation}
	\end{abet}
\end{proposition}
\begin{proof}
	To show assertion (a) we use Proposition \ref{prop:rho-restrict}(a) that for every $g \in \mathbb G^N$ it holds that $\rho^{-ij} (g) = \rho (g) + \rho (g+ij)$ if $ij \notin g$ and $\rho^{-ij} (g) =0$ if $ij \in g$. Hence,
	\begin{align*}
		\mathbb E (v, \rho^{-ij} ) & = \sum_{g \in \mathbb G^N} \rho^{-ij} (g) \cdot v(g) = \sum_{g \colon ij \notin g} \left( \, \rho (g) + \rho (g+ij) \, \right) v(g) \\[1ex]
		& = \sum_{g \colon ij \notin g} \left\{ \, \rho (g) v(g) + \rho (g+ij) v(g+ij) \, \right\} + \sum_{g \colon ij \notin g} \rho (g+ij) \left[ \, v(g) - v(g+ij) \, \right] \\[1ex]
		& = \sum_{g \in \mathbb G^N} \rho (g) \cdot v(g) + \sum_{g \colon ij \notin g} \rho (g+ij) \left[ \, v(g) - v(g+ij) \, \right] \\[1ex]
		& = \mathbb E(v, \rho ) - \sum_{g \colon ij \notin g} \rho (g+ij) \cdot \Delta_{ij} (v,g)
	\end{align*}
	To show assertion (b), we use Proposition \ref{prop:rho-restrict}(b) to derive that
	\begin{align*}
		\mathbb E (v, \rho^{-i} ) & = \sum_{g \colon i \notin N(g)} \left[ \, \rho(g) + \sum_{\varnothing \neq h \subseteq L_i} \rho (g \cup h) \, \right] v(g) \\[1ex]
		& = \sum_{g \colon i \notin N(g)} \rho(g) \cdot v(g) + \sum_{g \colon i \in N(g)} \rho(g) \cdot v(g \setminus L_i ) \\[1ex]
		& = \sum_{g \colon i \notin N(g)} \rho(g) \cdot v(g) + \sum_{g \colon i \in N(g)} \rho(g) \cdot v(g) + \sum_{g \colon i \in N(g)} \rho(g) \left[ \, v(g \setminus L_i ) - v(g) \, \right] \\[1ex]
		& = \mathbb E(v, \rho ) - \sum_{g \colon i \in N(g)} \rho (g) \cdot \Delta_i (v,g)
	\end{align*}
	This shows the assertion.
\end{proof}

\bigskip\noindent
The properties stated in Proposition \ref{prop-contributing} show that if links and players are contributing, their removal from a variable network game reduces the expected wealth that is generated. Furthermore, the removal or addition of so-called null players do not affect the expected wealth generated in variable network games, i.e., $\Delta_{i} (v,g)=0$ for some $g \in \mathbb G^N$ and $i \in N(g)$ implies that $\mathbb E(v, \rho^{-i}) = \mathbb E (v, \rho )$. Similarly, superfluous links do not affect wealth generation either in the sense that $\Delta_{ij} (v, g )=0$ for some $g \in \mathbb G^N$ and $ij \notin g$ implies that $\mathbb E(v, \rho^{-ij}) = \mathbb E (v, \rho )$.

\subsection{Allocation rules on variable network games}

The main objective of this paper is to investigate the allocation of the expected generated wealth in variable network games and the properties of the associated allocation rules. In particular, we consider natural probabilistic extensions of allocation rules from the class of network games to the class of variable network games. We focus hereby on the Myerson Value and the Position Value for network games. We show that the standard axiomatizations for both of these allocation rules extend to our framework of variable network games. 

An allocation rule on the class of (regular) network games is assigned to a network game as well as a certain given deterministic network, representing the interaction between the players. The allocation of the generated wealth is, therefore, conditioned on the particular network relationships between the players in the game.

On the other hand, variable network games are introduced as combinations of a network game and a network formation probability distribution. This implies that allocation rules should account for the stochastic nature of network formation processes and, consequently, the probabilistic relationships between the players. This is formalised in the next two definitions.

\begin{definition} \label{def:AllocationRule}
	An \textbf{allocation rule} on the class of variable network games is a mapping $\Psi \colon \mathbb V^N \times \mathbb P^N \to \mathbb R^N$ such that for every variable network game $(v, \rho ) \in \mathbb V^N \times \mathbb P^N$ it holds that $\Psi_i (v, \rho ) =0$ for every isolated player $i \in N_0 (g( \rho ))$.
\end{definition}

\noindent
The definition of an allocation rule is clearly a straightforward conceptual extension of the definition of allocation rules on the class of network games to the class of variable network games $\mathbb V^N \times \mathbb P^N$. In particular, any allocation rule on $\mathbb V^N$ can be extended to the larger class of variable network games $\mathbb V^N \times \mathbb P^N$ by taking its expected payoffs. This is formalised as follows.

\begin{definition} \label{def:Psi^Y}
	Let $Y \colon \mathbb V^N \times \mathbb G^N \to \mathbb R^N$ be an allocation rule on the class of network games. Then its \textbf{standard extension} to the class of variable network games is the allocation rule $\Psi^Y \colon \mathbb V^N \times \mathbb P^N \to \mathbb R^N$ defined by
	\begin{equation} \label{eq:Psi^Y}
		\Psi^Y (v, \rho ) = \sum_{g \in \mathbb G^N} \rho (g) \cdot Y(v,g) = \sum_{g \in \mathbb G (\rho )} \rho (g) \cdot Y (v,g)
	\end{equation}
	for every variable network game $(v, \rho ) \in \mathbb V^N \times \mathbb P^N$.\footnote{The implicitly stated property (\ref{eq:Psi^Y}) is a consequence of the fact that $\rho (g)=0$ if $g \in \mathbb G^N \setminus \mathbb G( \rho )$.}
\end{definition}

\noindent
The extension of an allocation rule simply assigns the expected payoff under that rule for the given network formation probability distribution. We establish in the subsequent sections that the standard extension of an allocation rule satisfies extensions of most of the properties of the original rule. 

\subsection{Balancedness properties of allocation rules}

The properties listed above for allocation rules on the class of network games can easily be extended to allocation rules on the class of variable network games. This is explored next for the properties introduced for network game allocation rules.

\paragraph{Balanced allocation rules}

An allocation rule $\Psi \colon \mathbb V^N \times \mathbb P^N \to \mathbb R^N$ is \emph{balanced} if for every variable network game $(v, \rho ) \in \mathbb V^N \times \mathbb P^N$ it holds that $\sum_{i \in N} \Psi_i (v, \rho ) = \mathbb E (v, \rho )$, i.e., the allocation rule exactly covers the expected wealth that is created in the variable network game.

\begin{proposition} \label{prop:Psi^Y balance}
	Let $Y \colon \mathbb V^N \times \mathbb G^N \to \mathbb R^N$ be a balanced allocation rule on the class of network games. Then its standard extension $\Psi^Y \colon \mathbb V^N \times \mathbb P^N \to \mathbb R^N$ is balanced.
\end{proposition}
\begin{proof}
	The proof of the assertion rests on the following:
	\begin{align*}
		\sum_{i \in N} \Psi^Y_i (v, \rho ) & = \sum_{i \in N} \sum_{g \in \mathbb G( \rho )} \rho (g) \cdot Y_i (v,g) = \sum_{g \in \mathbb G( \rho )} \rho (g) \cdot \left( \sum_{i \in N} Y_i (v,g) \, \right) \\[1ex]
		& = \sum_{g \in \mathbb G( \rho )} \rho (g) \cdot v(g) = \mathbb E (v, \rho )
	\end{align*}
	This indeed shows the assertion.
\end{proof}

\paragraph{Component Balanced allocation rules}

An allocation rule $\Psi \colon \mathbb V^N \times \mathbb P^N \to \mathbb R^N$ is \emph{component balanced} if for every variable network game $(v, \rho ) \in \mathbb V^N \times \mathbb P^N$ such that $v \colon \mathbb G^N \to \mathbb R$ is component additive, it holds that for every component $h \in C( g( \rho )) \colon$
\begin{equation}
	\sum_{i \in N(h)} \Psi_i (v, \rho ) = \sum_{g \in \mathbb G(\rho)} \rho (g) \cdot v (g \cap h)
\end{equation}
Component balance implies that all wealth that can be attributed to a certain component in the extent of the network formation probability distribution is allocated to the constituting members of that component. Hence, the wealth allocated through the rule exactly covers the expected wealth that is created in that component in the given variable network game.

\begin{proposition} \label{prop:Psi^Y component balance}
	Let $Y \colon \mathbb V^N \times \mathbb G^N \to \mathbb R^N$ be a component balanced allocation rule on the class of network games. Then its standard extension $\Psi^Y \colon \mathbb V^N \times \mathbb P^N \to \mathbb R^N$ is component balanced.
\end{proposition}
\begin{proof}
	Let $v \in \mathbb V^N$ be component additive and take any $\rho \in \mathbb P^N$. \\
	Next, take a component $h \in C(g (\rho ))$ and consider any network $g \subseteq g(\rho )$. Let $h' \in C(g)$ be a component of $g$ with $N(h) \cap N(h') \neq \varnothing$. Then, clearly, $h' \subseteq h$ and, therefore, from the component balance of $Y$ it then follows that
	\[
	\sum_{i \in N(h')} Y_i (v,g) =  v(h') .
	\]
	Hence, by component additivity of $v$, it follows that
	\[
		\sum_{i \in N(h)} Y_i (v,g) = \sum_{h' \in C(g \cap h)} \sum_{i \in N(h')} Y_i (v,g) = \sum_{h' \in C(g \cap h)} v(h') = v(g \cap h) .
	\]
	Therefore, we can conclude that
	\begin{align*}
		\sum_{i \in N(h)} \Psi^Y_i (v, \rho ) & = \sum_{i \in N(h)} \sum_{g \subseteq g(\rho )} \rho (g) \cdot Y_i (v,g)  \\[1ex]
		& = \sum_{g \subseteq g(\rho )} \rho (g) \cdot \left[ \, \sum_{i \in N(h)} Y_i (v,g) \, \right] = \sum_{g \subseteq g(\rho )} \rho (g) \cdot v(g \cap h)
	\end{align*}
	This completes the proof of the assertion.
\end{proof}

\subsection{The Expected Myerson and Position Values}

As discussed in Section 2.4, there are traditionally two principal allocation rules on the class of network games, namely the Myerson Value (\ref{eq:Y-m}) and the Position Value (\ref{eq:Y-p}). Both of these allocation rules can be extended to the class of variable network games through the consideration of their standard extension, reflecting the expected allocation of the generated wealth under these two values.

First, we explore extending the Myerson Value to the class of variable network games using the method explored above.

\begin{definition}
	The \textbf{Expected Myerson Value} is the allocation rule $\Psi^m \colon \mathbb V^N \times \mathbb P^N \to \mathbb R^N$ defined as the standard extension $\Psi^m = \Psi^{Y^m}$ of the Myerson Value $Y^m$ on the class of network games to the class of variable network games.
\end{definition}

\noindent 
As applied for the Myerson Value, we can also base an allocation rule on the class of variable network games on the formulated Position Value. This introduces the Expected Position Value.
\begin{definition}
The \textbf{Expected Position Value} is the allocation rule $\Psi ^{p} \colon \mathbb{V}^{N} \times \mathbb{P}^{N} \to  \mathbb{R}^{N}$ defined as the standard extension $\Psi ^{p}=\Psi ^{Y^{p}}$ of the Position Value $Y^{p}$ on the class of network games to the class of variable network games.
\end{definition}

\paragraph{Exploring expected values in the intermediated trade situation}

The Expected Myerson Value is simply the expectation of the Myerson payoff to a player that arises in every possible network that can form under the imposed network formation probability distribution. Similarly, the Expected Position Value is the expectation of the player's Position Value in the possible networks. We illustrate the computation of these two expected values by returning to the case of intermediated trade discussed in Examples \ref{Motiv-A} and \ref{Motiv-B}. 

\begin{example} \label{Motiv-C}
	Again consider the intermediated trade situation with player set $N = \{ S,B,I \}$, described in Example \ref{Motiv-A} and Figure 1. The generated wealth in this bilateral trade situation can be formulated as a network game $w \colon \mathbb G^N \to \mathbb R$ given by $w(g)=1$ if and only if $SB \in g$ and/or $\{ SI, BI \} \subseteq g$, and $w(g)=0$ otherwise. \\  Referring to Table 1, we denote by $\rho_1$ the network formation probability distribution  representing network formation under the hypothesis of independence of link formation with probabilities $q>0$ of $SI$ and $BI$ forming and $p>0$ of $SB$ forming. \\ For each resulting network we can thus compute the corresponding Myerson and Position Values. The following table collects the information for this case:
	
	\begin{table}[h!]
	\begin{center}
	\begin{tabular}{|c|c|c||c|c|}
		\hline
		$g$ & $w(g)$ & $\rho_1 (g)$ & $Y^m(w,g)$ & $Y^p (w,g)$ \\ \hline
		 $\varnothing$ & 0 & $(1-p)(1-q)^2$ & $(0,0,0)$ & $(0,0,0)$ \\
		 $\{ SI \}$ & 0 & $(1-p)(1-q)q$ & $(0,0,0)$ & $(0,0,0)$ \\
		 $\{ BI \}$ & 0 & $(1-p)(1-q)q$ & $(0,0,0)$ & $(0,0,0)$ \\
		 $\{ SB \}$ & 1 & $p (1-q)^2$ & $\left( \tfrac{1}{2}, \tfrac{1}{2},0 \right)$ & $\left( \tfrac{1}{2}, \tfrac{1}{2},0 \right)$ \\
		 $\{ SI,BI \}$ & 1 & $(1-p)q^2$ & $\left( \tfrac{1}{3}, \tfrac{1}{3} , \tfrac{1}{3} \right)$ & $\left( \tfrac{1}{4}, \tfrac{1}{4} , \tfrac{1}{2} \right)$ \\
		 $\{ SI,SB \}$ & 1 & $p(1-q)q$ & $\left( \tfrac{1}{2}, \tfrac{1}{2},0 \right)$ & $\left( \tfrac{1}{2}, \tfrac{1}{2},0 \right)$ \\
		 $\{ BI,SB \}$ & 1 & $p(1-q)q$ & $\left( \tfrac{1}{2}, \tfrac{1}{2},0 \right)$ & $\left( \tfrac{1}{2}, \tfrac{1}{2},0 \right)$ \\
		 $g_N$ & 1 & $p q^2$ & $\left( \tfrac{1}{2}, \tfrac{1}{2},0 \right)$ & $\left( \tfrac{5}{12}, \tfrac{5}{12} , \tfrac{1}{6} \right)$ \\
		\hline
	\end{tabular}
	\end{center}
	\caption{The Myerson and Position Values for Example \ref{Motiv-A}.}
	\end{table}
	
	\noindent
	From this we compute the Expected Myerson Value under independent link formation as
	\[
	\Psi^m_S (w,\rho_1) = \Psi^m_B (w,\rho_1) = \tfrac{1}{2} p + \tfrac{1}{3} (1-p)q^2 \qquad \mbox{and} \qquad \Psi^m_I (w,\rho_1) = \tfrac{1}{3} (1-p)q^2 .
	\]
	Furthermore, the Expected Position Value under independent link formation is
	\[
	\Psi^p_S (w,\rho_1) = \Psi^p_B (w,\rho_1) = \tfrac{1}{2} p + \tfrac{1}{4} q^2 - \tfrac{1}{3} pq^2 \qquad \mbox{and} \qquad \Psi^p_I (w,\rho_1) = \left( \tfrac{1}{2}  - \tfrac{1}{3} p \right) q^2 .
	\]
	Next consider as in Example \ref{Motiv-B} that the intermediary $I$ has a position of formalised leadership in the institutional setting and that a network $g$ is formable only if $I \in N(g)$. This is described by the probability distribution $\rho_2$ given in Table 2. This results in different Expected Myerson and Position Values, computed from Table 4.
	
	\begin{table}[h!]
	\begin{center}
	\begin{tabular}{|c|c||c|c|}
		\hline
		$g$ & $\rho_2 (g)$ & $Y^m(g,w)$ & $Y^p(g,w)$ \\ \hline
		 $\varnothing$ & $0$ & $(0,0,0)$ & $(0,0,0)$ \\
		 $\{ SI \}$ & $\frac{(1-p)(1-q)}{2-q}$ & $(0,0,0)$ & $(0,0,0)$ \\
		 $\{ BI \}$ & $\frac{(1-p)(1-q)}{2-q}$ & $(0,0,0)$ & $(0,0,0)$ \\
		 $\{ SB \}$ & $0$ & $\left( \tfrac{1}{2}, \tfrac{1}{2},0 \right)$ & $\left( \tfrac{1}{2}, \tfrac{1}{2},0 \right)$ \\
		 $\{ SI,BI \}$ & $\frac{(1-p)q}{2-q}$ & $\left( \tfrac{1}{3}, \tfrac{1}{3} , \tfrac{1}{3} \right)$ & $\left( \tfrac{1}{4}, \tfrac{1}{4} , \tfrac{1}{2} \right)$ \\
		 $\{ SI,SB \}$ & $\frac{p(1-q)}{2-q}$ & $\left( \tfrac{1}{2}, \tfrac{1}{2},0 \right)$ & $\left( \tfrac{1}{2}, \tfrac{1}{2},0 \right)$ \\
		 $\{ BI,SB \}$ & $\frac{p(1-q)}{2-q}$ & $\left( \tfrac{1}{2}, \tfrac{1}{2},0 \right)$ & $\left( \tfrac{1}{2}, \tfrac{1}{2},0 \right)$ \\
		 $g_N$ & $\frac{p q}{2-q}$ & $\left( \tfrac{1}{2}, \tfrac{1}{2},0 \right)$ & $\left( \tfrac{5}{12}, \tfrac{5}{12} , \tfrac{1}{6} \right)$ \\
		\hline
	\end{tabular}
	\end{center}
	\caption{The Myerson and Position Values for Example \ref{Motiv-B}.}
	\end{table}
	
	\noindent
	In this modified situation, we compute the Expected Myerson Values as
	\[
	\Psi^m_S (w,\rho_2) = \Psi^m_B (w,\rho_2) = \tfrac{1}{2} p + \tfrac{q(1-p)}{3(2-q)} \qquad \mbox{and} \qquad \Psi^m_I (w,\rho_2) = \tfrac{q(1-p)}{3(2-q)}
	\]
	From these computations, we conclude that for all values of $p,q \in [0,1]$ it holds that $\Psi^m_i (w,\rho_1) \leqslant \Psi^m_i (w,\rho_2)$ for every player $i \in \{ S,B,I \}$. Hence, the improved wealth generation situation due to institutional constraints on network formation translates to increased Expected Myerson payoffs to all constituting players in this intermediated trade situation.
	\\
	Furthermore, regarding the Expected Position Value we compute
	\[
	\Psi^p_S (w,\rho_2) = \Psi^p_B (w,\rho_2) = \tfrac{(3-10p)q}{12(2-q)} + \tfrac{p}{2-q} \qquad \mbox{and} \qquad \Psi^p_I (w,\rho_2) = \tfrac{(3-2p)q}{6(2-q)}
	\]
	Contrary to the Expected Myerson Value, for the Expected Position Value there is no unequivocal improvement of the expected payoffs under the introduced institutional restrictions on network formation. Indeed, the intermediary $I$'s Expected Position Value increases, but whether the Expected Position Value increases of $B$ and $S$, depends on the exact probabilities $p$ and $q$.
	\\
	To elaborate on this, we compare the Expected Position Values of the buyer $B$ and seller $S$ under two network formation rules. In particular, $\Psi^p_S (w, \rho_1) = \Psi^p_B (w, \rho_1) > \Psi^p_S (w, \rho_2) = \Psi^p_B (w, \rho_2)$ if and only if $(4p-3)(q-1)^2 q >0$, which is the case if and only if $\tfrac{3}{4} < p \leqslant 1$ and $0 < q < 1$. 
\end{example}

\section{Axiomatizing the Expected Myerson Value}

The Myerson Value on the class of network games can be fully characterized using standard axioms, in particular component balance and one other property, usually a fairness property---also denoted as the equal bargaining power property \citep{JacksonWolinsky1996}---and the balanced contributions property \citep{Slikker2007}. In this section we discuss similar characterizations of the Expected Myerson Value on the class of variable network games.

\paragraph{Equal Bargaining Power}

One of the main properties investigated in the literature since its inception by \citet{Myerson1977} is that of ``fairness'' in the allocation of wealth generated in a network. This refers to the idea that the removal of a single link from a network would affect both of its constituting players in equal measure. This has been referred to by \citet{JacksonWolinsky1996} as ``equal bargaining power'', which terminology we adopt here as well.

\begin{definition}
	Let $\Psi \colon \mathbb V^N \times \mathbb P^N \to \mathbb R^N$ be an allocation rule on the class of variable network games. Then $\Psi$ satisfies the \textbf{equal bargaining power property} if for every variable network game $(v, \rho ) \in \mathbb V^N \times \mathbb P^N$ and every link $ij \in g(\rho )$ in $\rho$'s extent it holds that
	\begin{equation}
		\Psi_i (v, \rho )- \Psi_i (v, \rho^{-ij} ) = \Psi_j (v, \rho )- \Psi_j (v, \rho^{-ij} )
	\end{equation}
\end{definition}

\noindent
The equal bargaining power property states that the removal of a single link affects the expected payoff to each of its constituting players equally. Note that the removal of a link can affect a player's payoff in a negative or a positive fashion. The formalisation of this property on the class of variable network games provides us with our first axiomatization of the Expected Myerson Value.

\begin{axiomatization}
	The Expected Myerson Value $\Psi^m$ is the unique allocation rule on the class of component additive variable network games that satisfies component balance as well as the equal bargaining power property.
\end{axiomatization}

\noindent
For a proof of this axiomatization we refer to Section 4.1.

\paragraph{Balanced Contributions}

\citet{Slikker2007} considered a different axiomatization of the Myerson Value on the class of network games. His approach is founded on the consideration of the effects of the removal of players from a network on the allocated values. Slikker proved an axiomatization based on the property that the effects of the removal of players are equalised. This is formalised on the class of variable network games as follows.

\begin{definition}
	Let $\Psi \colon \mathbb V^N \times \mathbb P^N \to \mathbb R^N$ be an allocation rule on the class of variable network games. Then $\Psi$ satisfies the \textbf{balanced contributions property} if for every variable network game $(v, \rho ) \in \mathbb V^N \times \mathbb P^N$ and all players $i,j \in N$ with $i \neq j$ it holds that
	\begin{equation}
		\Psi_i (v, \rho )- \Psi_i (v, \rho^{-j} ) = \Psi_j (v, \rho )- \Psi_j (v, \rho^{-i} )
	\end{equation}
\end{definition}

\noindent
Next we extend Slikker's axiomatization to the class of component additive variable network games founded on this definition of the balanced contributions property. This is stated as our second axiomatization of the Expected Myerson Value.

\begin{axiomatization}
	The Expected Myerson Value $\Psi^m$ is the unique allocation rule on the class of component additive variable network games that satisfies component balance as well as the balanced contributions property.
\end{axiomatization}

\noindent
For a proof of this second axiomatization we refer to Section 4.2.

\subsection{Proof of Axiomatization I}

We proceed with a proof of Axiomatization I in two subsequent steps. First, we show that the Expected Myerson Value indeed satisfies the properties of component balance and equal bargaining power. Second, we show that it is actually the unique allocation rule that satisfies these two properties.

\subsubsection*{$\Psi^m$ satisfies component balance}

We refer to \citet[Theorem 4]{JacksonWolinsky1996} for the fact that the Myerson Value $Y^m$ satisfies component balance on the class of component additive network games $\mathbb V^N$. Hence, by Proposition \ref{prop:Psi^Y component balance} it immediately follows that the Expected Myerson Value $\Psi^m = \Psi^{Y^m}$ satisfies component balance on the expanded class of component addirive variable network games $\mathbb V^N \times \mathbb P^N$. 

\subsubsection*{$\Psi^m$ satisfies the equal bargaining power property}

We can now show the following fundamental property that extends this insight to the framework of variable network games and the Expected Myerson Value.

\begin{lemma} \label{lem:Game-Psi^m}
	Let $(v, \rho ) \in \mathbb V^N \times \mathbb P^N$ be a variable network game. Then the Expected Myerson Value $\Psi^m (v, \rho ) = \phi \left(\omega_{(v, \rho )}\right)$ where $\omega_{(v, \rho )}$ is the cooperative game defined by
	\[
	\omega_{(v, \rho )} (S) = \sum_{g \in \mathbb G( \rho )} \rho (g) \cdot v(g|S) .
	\]
\end{lemma}
\begin{proof}
	Using the stated characterization by Jackson and Wolinsky, for every $(v, \rho ) \in \mathbb V^N \times \mathbb P^N$, we can write $\omega_{(v, \rho )} (S) = \sum_{g \in \mathbb G( \rho )} \rho (g) \cdot \hat\omega_{(v,g )} (S)$. By linearity of the Shapley Value \citep{Shapley1953} we now conclude that
	\[
	\phi \left( \omega_{(v, \rho )} \right) = \sum_{g \in \mathbb G( \rho )} \rho (g) \cdot \phi \left( \hat\omega_{(v,g)} \right) = \sum_{g \in \mathbb G( \rho )} \rho (g) \cdot Y^m (v,g) = \Psi^m (v, \rho ) .
	\]
	This shows the assertion.
\end{proof}

\bigskip\noindent
Next, to show the assertion that $\Psi^m$ satisfies the equal bargaining power property, let $(v, \rho ) \in \mathbb V^N \times \mathbb P^N$ be a variable network game. Furthermore, take any $i,j \in N$ with $i \neq j$, and define the derived cooperative game $\chi = \omega_{(v, \rho )} - \omega_{(v, \rho^{-ij} )}$.

\begin{claim}
	For every $S \subseteq N \setminus \{ i,j \} \colon \chi (S+i) = \chi (S+j) =0$.
\end{claim}

\noindent
Let $S \subseteq N \setminus \{ i,j \}$. To show the claim, consider that
\begin{align*}
\hspace*{-1em} \chi (S+i) & = \omega_{(v, \rho )} (S+i) - \omega_{(v, \rho^{-ij} )} (S+i) = \sum_{g \in \mathbb G^N} v(g| (S+i)) \cdot \left( \rho (g) - \rho^{-ij} (g) \, \right) \\[1ex]
	& = \sum_{g \in \mathbb G^N \colon ij \in g} v(g| (S+i)) \cdot \left( \rho (g) - \rho^{-ij} (g) \, \right) + \sum_{g \in \mathbb G^N \colon ij \notin g} v(g| (S+i)) \cdot \left( \rho (g) - \rho^{-ij} (g) \, \right) \\[1ex]
	& = \sum_{g \in \mathbb G^N \colon ij \in g} v(g| (S+i)) \cdot \left( \rho (g) - \rho^{-ij} (g) \, \right) - \sum_{g \in \mathbb G^N \colon ij \notin g} v(g| (S+i)) \cdot \rho (g+ij)
\end{align*}
By Proposition \ref{prop:rho-restrict}(a), $\rho^{-ij} (g) = \rho_{g_N-ij} (g) =0$ for all $g \in \mathbb G^N$ with $ij \in g$, we conclude that
\begin{align*}
	\chi (S+i) & = \sum_{g \in \mathbb G^N \colon ij \in g} v(g | (S+i)) \cdot \rho (g) - \sum_{g \in \mathbb G^N \colon ij \notin g} v(g| (S+i)) \cdot \rho (g+ij) \\[1ex]
	& = \sum_{g \in \mathbb G^N \colon ij \notin g} v((g+ij) | (S+i)) \cdot \rho (g+ij) - \sum_{g \in \mathbb G^N \colon ij \notin g} v(g| (S+i)) \cdot \rho (g+ij) \\[1ex]
	& = \sum_{g \in \mathbb G^N \colon ij \notin g} \rho (g+ij) \left( v((g+ij) | (S+i)) - v(g | (S+i)) \, \right)
\end{align*}
However, if $ij \notin g$ we have that $(g+ij) | (S+i) = g| (S+i)$ since $j \notin S$. Hence, $v((g+ij) | (S+i)) = v(g | (S+i))$ and, therefore, $\chi (S+i)=0$. Similarly, we also conclude that $\chi (S+j)=0$.  This shows Claim A.

\begin{claim}
	$\phi_i \left( \omega_{(v, \rho )} \right) - \phi_i \left( \omega_{(v, \rho^{-ij} )} \right) = \phi_j \left( \omega_{(v, \rho )} \right) - \phi_j \left( \omega_{(v, \rho^{-ij} )} \right)$.
\end{claim}

\noindent
First, note that both players $i$ and $j$ are symmetric in the derived game $\chi$. From the previous claim, it follows by application of the symmetry axiom to the Shapley Value that $\phi_i ( \chi ) = \phi_j (\chi )$. Hence, the assertion of Claim B follows immediately from the additivity axiom as satisfied by the Shapley Value. (See Section 2.1 for the required definitions.)

\subsubsection*{There is at most one allocation rule that satisfies component balance as well as the equal bargaining power property}

To prove uniqueness of an allocation rule satisfying both stated properties, assume by contradiction that there exist two such allocation rules $\Psi^1 \neq \Psi^2$ that satisfy component balance as well as the equal bargaining power property.

Let $v \in \mathbb V^N$ be a given network game. Now consider a network formation probability distribution $\tilde \rho \in \mathbb P^N$ such that $\Psi^1 (v, \tilde \rho ) \neq \Psi^2 (v, \tilde \rho )$ and for every $\rho \in \mathbb P^N$ with $\# g( \rho ) < \# g( \tilde \rho)$ it holds that $\Psi^1 (v, \rho ) = \Psi^2 (v, \rho )$. Hence, the size of the extent $\# g( \tilde \rho )$ is minimal regarding the property that both of these allocation rules are unequal.

Note that for $\rho_0 \in \mathbb P^N$ with $\rho_0 (g_0)=1$ and $\rho_0 (g)=0$ for $g \neq g_0$, where $g_0 = \varnothing$. Then $\Psi^1 (v, \rho_0 ) = \Psi^2 (v, \rho_0 ) =0$. Hence, there exists some $\tilde \rho \neq \rho_0$ as described.

\medskip\noindent
Next, let $i,j \in N$ with $i \neq j$ with $ij \in g ( \tilde \rho)$.  Then by Proposition \ref{prop:rho-restrict}(a) it follows that $g (\tilde \rho^{-ij} ) = g (\tilde \rho) -ij$ and by the definition of $\tilde \rho$ we conclude that $\Psi^1 (v, \tilde \rho^{-ij} ) = \Psi^2 (v, \tilde \rho^{-ij} )$. Therefore, by the equal bargaining power property for both allocation rules,
\begin{align*}
	\Psi^1_i (v, \tilde \rho ) - \Psi^1_j (v, \tilde \rho) & = \Psi^1_i \left( v, \tilde\rho^{-ij} \right) - \Psi^1_j \left( v, \tilde\rho^{-ij} \right) \\[1ex]
	& = \Psi^2_i \left( v, \tilde\rho^{-ij} \right) - \Psi^2_j \left( v, \tilde\rho^{-ij} \right) = \Psi^2_i (v, \tilde \rho ) - \Psi^2_j (v, \tilde \rho)
\end{align*}
Hence, we have shown that
\[
\Psi^1_i (v, \tilde \rho ) - \Psi^2_i (v, \tilde \rho) = \Psi^1_j (v, \tilde \rho ) - \Psi^2_j (v, \tilde \rho)
\]
In particular, we conclude now that there exist numbers $\{ \xi_h \in \mathbb R \mid h \in C(g( \tilde \rho)) \}$ for which it holds that $\Psi^1_i (v, \tilde \rho ) - \Psi^2_i (v, \tilde \rho) = \xi_h$ for all $i \in N(h)$. Moreover, by component balance of the two allocation rules and the component additivity of $v$, for every $h \in C(g( \tilde \rho )) \colon$
\[
\sum_{i \in N(h)} \Psi^1_i (v, \tilde \rho ) = \sum_{i \in N(h)} \Psi^2_i (v, \tilde \rho ) = \sum_{g \in \mathbb G(\rho)} \rho (g) \cdot v (g \cap h) .
\]
Hence, we have that
\[
0=\sum_{i \in N(h)} \Psi^1_i (v, \tilde \rho ) - \sum_{i \in N(h)} \Psi^2_i (v, \tilde \rho ) = \sum_{i \in N(h)} \left( \Psi^1_i (v, \tilde \rho ) - \Psi^2_i (v, \tilde \rho ) \, \right) = \# N(h) \cdot \xi_h ,
\]
implying that $\xi_h =0$. Therefore, $\Psi^1_i (v, \tilde \rho ) = \Psi^2_i (v, \tilde \rho )$ for all $i \in N$.

This shows that, indeed, there is at most one allocation rule that satisfies both component balance as well as the equal bargaining power property. Clearly, this shows that the Expected Myerson Value is the unique allocation rule satisfying both of these properties. This completes the proof of Axiomatization I.

\subsection{Proof of Axiomatization II}

In the proof of Axiomatization I we already have shown that the Expected Myerson Value satisfies component balance. Therefore, it remains to be shown that it also satisfies the balanced contributions property and that it is the unique allocation rule on the class of variable network games to satisfy both of these properties. 

\subsubsection*{$\Psi^m$ satisfies the balanced contributions property}

Let $(v, \rho ) \in \mathbb V^N \times \mathbb P^N$ be a variable network game. Furthermore, let $i,j \in N$ with $i \neq j$. Then
\begin{align*}
	\Psi^m_i (v, \rho ) - \Psi^m_i (v, \rho^{-j} ) & = \sum_{g \in \mathbb G^N} Y^m_i (v,g) \cdot \rho (g) - \sum_{g \in \mathbb G^N} Y^m_i (v,g) \cdot \rho^{-j} (g) \\[1ex]
	& = \sum_{g \subseteq g (\rho )} Y^m_i (v,g) \cdot \rho (g) - \sum_{g \subseteq g (\rho^{-j})} Y^m_i (v,g) \cdot \rho^{-j} (g)
\end{align*}
By Proposition \ref{prop:rho-restrict}(b) it holds that $g( \rho^{-j}) = g (\rho ) - L_j (g ( \rho ))$. Hence,
\[
\Psi^m_i (v, \rho ) - \Psi^m_i (v, \rho^{-j} ) = \sum_{\scriptsize \begin{array}{c} g \subseteq g (\rho ) - L_j (g( \rho )) \\ h \subseteq L_j (g ( \rho )) \end{array}} Y^m_i (v,g+h) \cdot \rho (g+h) - \sum_{g \subseteq g (\rho ) - L_j (g ( \rho ))} Y^m_i (v,g) \cdot \rho^{-j} (g)
\]
Using Proposition \ref{prop:rho-restrict}(b), we conclude that
\begin{align*}
	\Psi^m_i (v, \rho ) - \Psi^m_i (v, \rho^{-j} ) & = \sum_{\scriptsize \begin{array}{c} g \subseteq g (\rho ) - L_j (g( \rho )) \\ h \subseteq L_j (g ( \rho )) \end{array}} Y^m_i (v,g+h) \cdot \rho (g+h) \\[1ex]
	& \hspace{4em} - \sum_{g \subseteq g (\rho ) - L_j (g ( \rho ))} Y^m_i (v,g) \cdot \left( \rho (g) + \sum_{\varnothing \neq h' \subseteq L_j ( g ( \rho ))} \rho (g+h') \right) \\[1ex]
	& = \sum_{\scriptsize \begin{array}{c} g \subseteq g (\rho ) - L_j (g( \rho )) \\ h \subseteq L_j (g ( \rho )) \end{array}} Y^m_i (v,g+h) \cdot \rho (g+h) \\[1ex]
	& \hspace{4em} - \sum_{g \subseteq g (\rho ) - L_j (g ( \rho ))} Y^m_i (v,g) \cdot \rho (g) \\[1ex]
	& \hspace{4em} - \sum_{\scriptsize \begin{array}{c} g \subseteq g (\rho ) - L_j (g( \rho )) \\ \varnothing \neq h' \subseteq L_j (g ( \rho )) \end{array}} Y^m_i (v,g) \cdot \rho (g+h') \\[1ex]
	& = \sum_{\scriptsize \begin{array}{c} g \subseteq g (\rho ) - L_j (g( \rho )) \\ \varnothing \neq h \subseteq L_j (g ( \rho )) \end{array}} \left( Y^m_i (v,g+h) - Y^m_i (v,g) \right) \cdot \rho (g+h) \\[1ex]
	& = \sum_{g \subseteq g(\rho ) \colon L_j (g) \neq \varnothing} \left( Y^m_i (v,g) - Y^m_i (v,g - L_j (g)) \right) \cdot \rho (g)
\end{align*}
Next, consider the networks $g \subseteq g ( \rho )$ with $L_j (g) = \varnothing$. Obviously, $Y^m_i (v,g) - Y^m_i (v,g - L_j (g))=0$ for these networks.  These networks can therefore be added to the expression derived above. Therefore, we can rewrite
\[
\Psi^m_i (v, \rho ) - \Psi^m_i (v, \rho^{-j} ) = \sum_{g \subseteq g(\rho )} \left( Y^m_i (v,g) - Y^m_i (v,g - L_j (g)) \right) \cdot \rho (g)
\]
In a similar way, we can derive that
\[
\Psi^m_j (v, \rho ) - \Psi^m_j (v, \rho^{-i} ) = \sum_{g \subseteq g(\rho )} \left( Y^m_j (v,g) - Y^m_j (v,g - L_i (g)) \right) \cdot \rho (g)
\]
From the balanced contributions property for the Myerson Value $Y^m$ on the class of network games, we also know that for every $g \in \mathbb G^N \colon$
\[
Y^m_i (v,g) - Y^m_i (v,g - L_j (g)) = Y^m_j (v,g) - Y^m_j (v,g - L_i (g))
\]
We conclude that $\Psi^m$ indeed satisfies the balanced contributions property on the class of variable network games.

\subsubsection*{There is at most one allocation rule that satisfies component balance as well as the balanced contributions property}

To show the uniqueness of an allocation rule that satisfies component balance as well as the balanced contributions property, we use an induction argument on the number of links making up $\rho$'s extent $g( \rho )$.

Suppose that $\Psi^1$ and $\Psi^2$ are two allocation rules on the class of variable network games that both satisfy component balance as well as the balanced contributions property.

Let $(v, \rho ) \in \mathbb V^N \times \mathbb P^N$ be some given variable network game with $\# g ( \rho )=1$---indicating that the extent of $\rho$ consists of two connected players through a single link. Then the balanced contributions property is fully equivalent to the equal bargaining power property. Hence, by Axiomatization I $\Psi^1 (v, \rho ) = \Psi^2 (v, \rho ) = \Psi^m (v, \rho )$, showing the uniqueness of the allocation rule for this case.

Let $k \in \mathbb N$. Assume the induction hypothesis that for every variable network game $(v, \rho ) \in \mathbb V^N \times \mathbb P^N$ with $\# g ( \rho ) \leqslant k$ it holds that $\Psi^1 (v, \rho ) = \Psi^2 (v, \rho ) = \Psi^m (v, \rho )$.

Let $(v, \rho ) \in \mathbb V^N \times \mathbb P^N$ be some given variable network game with $\# g ( \rho )=k+1$. Take $h \in C(g ( \rho ))$ and let $i,j \in N(h)$. Then clearly by definition $\# g( \rho^{-i}) \leqslant k$ as well as $\# g( \rho^{-j}) \leqslant k$. By the induction hypothesis, $\Psi^1 (v, \rho^{-i} ) = \Psi^2 (v, \rho^{-i} )$ as well as $\Psi^1 (v, \rho^{-j} ) = \Psi^2 (v, \rho^{-j} )$.

Furthermore, by the balanced contributions property for $\Psi^1$ and $\Psi^2$ it then holds that
\begin{align*}
	\Psi^1_i (v, \rho ) - \Psi^1_j (v, \rho ) & = \Psi^1_i (v, \rho^{-j} ) - \Psi^1_j (v, \rho^{-i} ) \\[1ex]
	& =  \Psi^2_i (v, \rho^{-j} ) - \Psi^2_j (v, \rho^{-i} ) = \Psi^2_i (v, \rho ) - \Psi^2_j (v, \rho ) .
\end{align*}
 Hence, for any $h \in C(g( \rho ))$ there exists some $\xi_h \in \mathbb R$ such that for all $i,j \in N(h) \colon$
 \[
 \Psi^1_i (v, \rho ) - \Psi^2_i (v, \rho ) = \Psi^1_j (v, \rho ) - \Psi^2_j (v, \rho ) = \xi_h .
 \]
 Moreover, by component balance of both $\Psi^1$ and $\Psi^2$ it holds that
 \[
 \sum_{i \in N(h)} \Psi^1_i (v, \rho ) = \sum_{i \in N(h)} \Psi^2_i (v, \rho ) = \sum_{g \in \mathbb G(\rho)} \rho (g) \cdot v (g \cap h)
 \]
 This implies that
 \[
 0 = \sum_{i \in N(h)} \left( \, \Psi^1_i (v, \rho ) - \Psi^2_i (v, \rho ) \, \right) =\# N(h) \times \xi_h .
 \]
 It follows, therefore, that $\xi_h=0$ and, hence, $\Psi^1_i (v, \rho ) = \Psi^2_i (v, \rho )$ for all $i \in N(h)$. By extension, $\Psi^1 (v, \rho ) = \Psi^2 (v, \rho ) = \Psi^m (v, \rho )$, showing the assertion and, consequently, Axiomatization II.
 
 \section{An axiomatization of the Expected Position Value}
 
 Next we consider axiomatizing the Expected Position Value along the lines of the axiomatization developed by \citet{Slikker2007} for network games. With regard to axiomatizations of the Position Value for probabilistic settings, we note that \citet{Ghintran2012} has done this for probabilistic communication situations and that \citet{ProbValue2021} develop an axiomatization for probabilistic network games. Here we pursue an extension of these axiomatizations to the class of variable network games.
 
 \paragraph{Balanced Link Contributions}
 
 We first extend the balanced link contributions property to the class of variable network games and, subsequently, extend Slikker's axiomatization founded on this property to variable network games. We can quite straightforwardly extend the balanced link contributions property from the class of network games to the class of variable network games.
 
 \begin{definition}
Let $\Psi  \colon \mathbb{V}^{N} \times \mathbb{P}^{N} \to \mathbb{R}^{N}$ be an allocation rule on the class of variable network games. Then $\Psi $ satisfies the \textbf{balanced link contributions property} if for every variable network game $(v,p) \in \mathbb{V}^{N} \times \mathbb{P}^{N}$ and all players $i,j \in N$ with $i \neq j$, it holds that 
\begin{equation}
\sum_{jk \in L_{j} (g(\rho ))} \left[ \Psi _{i} (v,\rho )-\Psi_{i} \left( v,\rho ^{-jk}\right) \, \right] = \sum_{ik\in L_{i}(g(\rho ))} \left[ \Psi _{j} (v,\rho )-\Psi _{j} \left( v,\rho^{-ik} \right) \, \right] .
\end{equation}
\end{definition}

\noindent
The balanced link contributions property implies that the sum of changes to $i$'s payoff when $j$'s links are removed one at a time probabilistically from the extent of $\rho $ is the same as the sum of changes to $j$'s payoffs when $i$'s links are removed probabilistically one at a time. 

Next we extend Slikker's axiomatization \citep{Slikker2007} to the class of variable network games founded on this definition of the balanced link contributions property.

\begin{axiomatization}
	The Expected Position Value $\Psi ^{p}$ is the unique allocation rule on the class of component additive variable network games that satisfies component balance as well as the balanced link contributions property.
\end{axiomatization}

\subsection{Proof of Axiomatization III}

We proceed with a proof of Axiomatization III by first showing that the Expected Position Value indeed satisfies the properties of component balance and balanced link contributions property. Second, we show that it is actually the unique allocation rule that satisfies these two properties.

\subsubsection*{$\Psi ^{p}$ satisfies component balance}

We refer to Slikker (2007, Theorem $3.1$) for the fact that the Position Value $Y^{p}$ satisfies component balance on the class of network games $\mathbb{V}^{N}$. Hence, by Proposition \ref{prop:Psi^Y component balance}, it immediately follows that the Expected Position Value $\Psi ^{p}=\Psi ^{Y^{p}}$ is component balanced on the expanded class of variable network games $\mathbb{V}^{N}\times \mathbb{P}^{N}$.

\subsubsection*{$\Psi ^{p}$ satisfies the balanced link contributions property}

Let $(v,\rho )\in \mathbb{V}^{N}\times \mathbb{P}^{N}$ be a variable network game. Furthermore, let $i,j\in N$ with $i\neq j$. Then 
\begin{align*}
\sum_{jk\in L_{j}(g(\rho ))} & \left[ \Psi _{i}^{p}(v,\rho ) -\Psi_{i}^{p}\left( v,\rho ^{-jk}\right) \, \right] = \\
& = \sum_{jk\in L_{j}(g(\rho ))} \left[ \sum_{g\subseteq g(\rho )} Y_{i}^{p}(v,g)\cdot \rho (g)-\sum_{g\subseteq g(\rho ^{-jk})} Y_{i}^{p}(v,g)\cdot \rho ^{-jk}(g) \right] \\[2ex]
& = \sum_{jk\in L_{j}(g(\rho ))} \left[ \sum_{g\subseteq g(\rho ) \colon jk\in g} Y_{i}^{p}(v,g)\cdot \rho (g)+\sum_{g\subseteq g(\rho ) \colon jk\notin g} Y_{i}^{p}(v,g)\left( \rho (g)-\rho ^{-jk}(g)\right) \right]
\end{align*}
given that $g(\rho ^{-jk})= g(\rho )-jk$.

Furthermore, by Proposition \ref{prop:rho-restrict}(a), if $jk\in g$, then $\rho ^{-jk}(g)=0$. On the other hand, if $jk\notin g$, it holds that $\rho ^{-jk}(g)=\rho (g)+\rho (g+jk)$. Hence, $\rho (g)-\rho ^{-jk}(g)=-\rho (g+jk)$. Therefore, 
\begin{align*}
\sum_{jk\in L_{j}(g(\rho ))} & \left[ \Psi _{i}^{p}(v,\rho )-\Psi_{i}^{p}\left( v,\rho ^{-jk}\right) \right] \\
&= \sum_{jk\in L_{j}(g(\rho ))} \left[ \sum_{g\subseteq g(\rho ) \colon jk\in g} Y_{i}^{p}(v,g)\cdot \rho (g)-\sum_{g\subseteq g(\rho ) \colon jk\notin g} Y_{i}^{p}(v,g)\left( \rho (g+jk)\right) \right] .
\end{align*}
Note that $\{g\subseteq g(p)|jk\in g\} = \{g+jk|g\subseteq g(p)$ and $jk\notin g\}$. This allows us to conclude that 
\begin{align*}
\sum_{jk\in L_{j}(g(\rho ))} & \left[ \Psi _{i}^{p}(v,\rho )-\Psi_{i}^{p}\left( v,\rho ^{-jk}\right) \right] = \\
&= \sum_{jk\in L_{j}(g(\rho ))} \left[ \underset{g\subseteq g(\rho ) \colon jk\notin g}\sum Y_{i}^{p}(v,g+jk)\cdot \rho (g+jk)-\sum_{g\subseteq g(\rho ) \colon jk\notin g} Y_{i}^{p}(v,g)\cdot \rho (g+jk)\right] \\[2ex]
&= \sum_{jk\in L_{j}(g(\rho ))} \left[ \sum_{g\subseteq g(\rho ) \colon jk\notin g} \rho (g+jk)\left( Y_{i}^{p}(v,g+jk)-Y_{i}^{p}(v,g)\right) \right] \\[2ex]
&= \sum_{jk\in L_{j}(g(\rho ))} \left[ \sum_{g\subseteq g(\rho ) \colon jk\in g} \rho (g)\left( Y_{i}^{p}(v,g)-Y_{i}^{p}(v,g-jk)\right) \right] \\[2ex]
 &= \sum_{g\subseteq g(\rho )}  \sum_{jk\in L_{j}(g)} \rho (g)\left( Y_{i}^{p}(v,g)-Y_{i}^{p}(v,g-jk)\right) \\[2ex]
&= \sum_{g\subseteq g(\rho )} \rho (g)\text{ }\left[ \sum_{jk\in L_{j}(g)} \left( Y_{i}^{p}(v,g)-Y_{i}^{p}(v,g-jk)\right) \right]
\end{align*}
Similarly, we conclude that
\begin{equation*}
\sum_{ik\in L_{i}(g(\rho ))} \left[ \Psi _{j}^{p}(v,\rho )-\Psi_{j}^{p}\left( v,\rho ^{-ik}\right) \right] =\sum_{g\subseteq g(\rho )} \rho (g) \, \left[ \sum_{ik\in L_{i}(g)} \left( Y_{j}^{p}(v,g)-Y_{j}^{p}(v,g-ik)\right) \right] .
\end{equation*}
By the balanced link contribution property of the Position Value for network games, we know that for all networks $g\in \mathbb{G}^{N}$ it holds that
\begin{equation*}
\sum_{jk\in L_{j}(g)} \left( Y_{i}^{p}(v,g)-Y_{i}^{p}(v,g-jk)\right) =\sum_{ik\in L_{i}(g)} \left( Y_{j}^{p}(v,g)-Y_{j}^{p}(v,g-ik)\right)
\end{equation*}
implying that 
\begin{equation*}
\sum_{jk\in L_{j}(g(\rho ))} \left[ \Psi _{i}^{p}(v,\rho )-\Psi_{i}^{p}\left( v,\rho ^{-jk}\right) \right] =\sum_{ik\in L_{i}(g(\rho ))} \left[ \Psi _{j}^{p}(v,\rho )-\Psi _{j}^{p}\left( v,\rho^{-ik}\right) \right] .
\end{equation*}
This shows the assertion that the Expected Position Value indeed satisfies the balanced link contributions property.

\subsubsection*{There is at most one allocation rule that satisfies component balance as well as the balanced link contributions property}

The proof closely follows the original characterization of \citet{Slikker2007}.

Let $\rho \in \mathbb{P}^{N}$ be fixed. Suppose $\Psi  \colon \mathbb{V}^{N}\times \mathbb{P}^{N} \to \mathbb{R}^{N}$ satisfies component balance and the balanced link contributions property. We intend to show that $\Psi =\Psi ^{p}$. The proof is by induction on $\#g(\rho )$, the number of links in the extent of $\rho $.

First, consider $\# g( \rho )=0$. Then obviously $\rho =\rho _{0}$ given by $\rho_0 (g_0) =1$ and $\rho _{0}(g)=0$ for all networks $g\neq g_0$, where $g_0= \varnothing$. As a consequence, $\Psi (\cdot ,\rho )=\Psi ^{p}(\cdot ,\rho )=0$ by Definition \ref{def:AllocationRule} of an allocation rule on the class of variable network games.

Next, let $k\in \mathbb{N}$. By the induction hypothesis, assume that $\Psi (\cdot ,\rho ^{\prime })=\Psi ^{p}(\cdot ,\rho ^{\prime })$ for all $\rho ^{\prime} \in \mathbb{P}^{N}$ with $\# g(\rho ^{\prime })\leqslant k-1$ (with $k-1\geqslant 0$).

Take $v\in \mathbb{V}^{N}$ and let $\rho \in \mathbb{P}^{N}$ with $\# g(\rho )=k$. Take any component $h\in C(g(\rho ))$. We may assume, without loss of generality, that $N(h)= \{ 1,2,\ldots ,m \} \subseteq N$. By the balanced link contributions property, for every $j\in N(h)$, $j\neq 1$, it holds that
\begin{equation*}
\sum_{jk\in L_{j}(g(\rho ))} \left[ \Psi _{1}(v,\rho )-\Psi_{1}\left( v,\rho ^{-jk}\right) \right] =\sum_{1k\in L_{1}(g(\rho ))}\left[ \Psi _{j}(v,\rho )-\Psi _{j}\left( v,\rho ^{-1k}\right) \right]
\end{equation*}
This implies that
\begin{equation*}
\#L_{j}(g(\rho )) \cdot \Psi _{1}(v,\rho )-\#L_{1}(g(\rho )) \cdot \Psi _{j}(v,\rho )= \sum_{jk\in L_{j}(g(\rho ))} \Psi _{1} \left( v,\rho ^{-jk}\right) - \sum_{1k\in L_{1}(g(\rho ))} \Psi _{j}\left( v,\rho ^{-1k}\right) .
\end{equation*}
Now by the induction hypothesis, we conclude that
\begin{equation}
\sum_{jk\in L_{j}(g(\rho ))} \Psi _{1}\left( v,\rho ^{-jk}\right) = \sum_{jk\in L_{j}(g(\rho ))} \Psi _{1}^{p}\left( v,\rho^{-jk}\right) 
\end{equation}
as well as 
\begin{equation}
\sum_{1k\in L_{1}(g(\rho ))} \Psi _{j}\left( v,\rho ^{-1k}\right) = \sum_{1k\in L_{1}(g(\rho ))} \Psi _{j}^{p}\left( v,\rho^{-1k}\right) .
\end{equation}
Therefore, we deduce that 
\begin{equation*}
\#L_{j}(g(\rho )) \cdot \Psi _{1}(v,\rho )-\#L_{1}(g(\rho )) \cdot \Psi _{j}(v,\rho )= \sum_{jk\in L_{j}(g(\rho ))} \Psi _{1}^{p}\left( v,\rho^{-jk}\right) -\sum_{1k\in L_{1}(g(\rho ))} \Psi _{j}^{p}\left( v,\rho ^{-1k}\right) .
\end{equation*}
Furthermore, by component balance and component additivity of $v$, we have 
\begin{equation*}
\sum_{i=1}^m  \Psi _{i}(v,\rho )=\sum_{g\in G(\rho )} \rho (g)\cdot v(g\cap h) \text{.}
\end{equation*}
Hence, we have a system of $m$ equations in $m$ unknowns. It is a straightforward exercise to show that this is a regular system in $m$ variables $\Psi _{1}(v,\rho ),\Psi _{2}(v,\rho ),\ldots ,\Psi _{m}(v,\rho )$. Consequently, it has a unique solution.

Since the Position Value satisfies balanced link contributions and component balance, it follows that $\Psi _{1}^{p}(v,\rho ),\Psi _{2}^{p}(v,\rho ),\ldots ,\Psi _{m}^{p}(v,\rho )$ is a solution and, hence, it has to be the unique solution. We conclude that $\Psi (v,\rho )=\Psi^{p}(v,\rho )$ for any $\rho \in \mathbb{P}^{N}$ with $\# g(\rho )=k$.

This completes the proof of Axiomatization III.

\section{Concluding remarks}

In this paper we have shown that the seminal Myerson Value and Position Value axiomatizations naturally extend to broader classes of network-based wealth generating settings. We consider the broad class of probabilistic network-based wealth generation situations based on general network-based probabilities rather than independent link-based probabilities considered thus far in the literature. furthermore, we have investigated value generating situations based on network games \citep{JacksonWolinsky1996} rather than on the more restrictive class of communication situations \citep{Myerson1977,Myerson1980}. 

For future directions of research, we suggest the investigation of alternative axiomatizations of the Expected Myerson Value and Expected Position Value and the exploration of values outside the class of expected values for variable network games. Furthermore, one may explore whether the Expected Myerson Value and the Expected Position Value can be represented as the marginals of some potential function in the sense of \citet{HartMascolell1989} extended to the class of variable network games.

\singlespace
\bibliographystyle{ecta}
\bibliography{RPDB}

\end{document}